\documentclass[twocolumn,aps,prb,10pt,superscriptaddress]{revtex4-2}

\usepackage{amsmath}
\usepackage{amssymb}
\usepackage{microtype}
\usepackage{graphicx}
\usepackage{hyperref}
\usepackage{dsfont}
\usepackage{newtxtext}
\usepackage[varvw]{newtxmath}
\usepackage{inconsolata}
\usepackage{slashed}
\usepackage{bbm}
\usepackage{tabularx}

\hypersetup{
    colorlinks,
    linkcolor={blue},
    citecolor={blue},
    urlcolor={blue}
}

\graphicspath{{./}{./figs/}{../figs/}}

\newcommand{\rme}{\mathrm{e}}

\newcommand{\rmd}{\mathrm{d}}

\newcommand{\idCliff}{\mathds{1}_{2N/3}}

\usepackage[dvipsnames]{xcolor}
\usepackage[normalem]{ulem}

\begin{document}

\title{%
Fractionalized quantum criticality in spin-orbital liquids from field theory beyond the leading order
}

\begin{flushright}
      \normalsize LTH 1253
\end{flushright}
\vspace{-1.2cm}

\author{Shouryya Ray}
\affiliation{Institut f\"ur Theoretische Physik and W\"urzburg-Dresden Cluster of Excellence ct.qmat, TU Dresden, 01062 Dresden, Germany}

\author{Bernhard Ihrig}
\affiliation{Institute for Theoretical Physics, University of Cologne, 50937 Cologne, Germany}

\author{Daniel Kruti}
\affiliation{Institute for Theoretical Physics, University of Cologne, 50937 Cologne, Germany}

\author{John A.\ Gracey}
\affiliation{Theoretical Physics Division, Department of Mathematical Sciences, University of Liverpool, Liverpool, L69 3BX, United Kingdom}

\author{Michael~M.~Scherer}
\affiliation{Institute for Theoretical Physics, University of Cologne, 50937 Cologne, Germany}

\author{Lukas Janssen}
\affiliation{Institut f\"ur Theoretische Physik and W\"urzburg-Dresden Cluster of Excellence ct.qmat, TU Dresden, 01062 Dresden, Germany}

\begin{abstract}
Two-dimensional spin-orbital magnets with strong exchange frustration have recently been predicted to facilitate the realization of a quantum critical point in the Gross-Neveu-SO(3) universality class.
In contrast to previously known Gross-Neveu-type universality classes, this quantum critical point separates a Dirac semimetal and a long-range-ordered phase, in which the fermion spectrum is only partially gapped out.
Here, we characterize the quantum critical behavior of the Gross-Neveu-SO(3) universality class by employing three complementary field-theoretical techniques beyond their leading orders.
We compute the correlation-length exponent $\nu$, the order-parameter anomalous dimension $\eta_\phi$, and the fermion anomalous dimension $\eta_\psi$ using a three-loop $\epsilon$ expansion around the upper critical space-time dimension of four, a second-order large-$N$ expansion (with the fermion anomalous dimension obtained even at the third order), as well as a functional renormalization group approach in the improved local potential approximation.
For the physically relevant case of $N=3$ flavors of two-component Dirac fermions in 2+1 space-time dimensions, we obtain the estimates $1/\nu = 1.03(15)$, $\eta_\phi = 0.42(7)$, and $\eta_\psi = 0.180(10)$ from averaging over the results of the different techniques, {with the displayed uncertainty representing the degree of consistency among the three methods}.
\end{abstract}

\date{\today}

\maketitle

\section{Introduction}

A quantum critical point is a continuous phase transition at absolute zero temperature, driven by some nonthermal parameter, such as pressure, doping, or magnetic field.
In many cases, such a transition is characterized by fluctuations of a local order parameter alone. 
The behavior of the system near criticality can then be understood via the quantum-to-classical mapping, which relates the universal properties of the quantum transition in $d$ spatial dimensions to those of a corresponding thermal transition in $d+z$ dimensions. 
Here, the dynamical critical exponent $z$ corresponds to the relative scaling of the correlation time and the correlation length near criticality \cite{sachdevbook}.
In the search for transitions beyond this quantum-to-classical paradigm, quantum critical points that are characterized not only by order-parameter fluctuations, but in addition feature gapless fermion degrees of freedom, occupy center stage \cite{*[] [{; and references therein.}] boyack20}.
The presence of low-energy fermion fluctuations at such a transition prevents a naive mapping to a corresponding classical critical point. 
Such a fermionic quantum critical point therefore usually realizes a novel quantum universality class, characterized by a set of universal exponents that significantly differs from those of the usual bosonic universality classes.

In this context, the (2+1)-dimensional Gross-Neveu-type universality classes have received significant attention in recent years~%
\cite{%
mihaila17,zerf17,ihrig18,janssen18,
gracey16,gracey17,gracey18,
iliesiu16,iliesiu18,
braun11,janssen14,vacca15,knorr16,gies17,knorr18,dabelow19,yin20%
}.
They describe transitions between a symmetric Dirac semimetal phase, in which the Fermi surface consists of isolated linear band-crossing points, and a long-range ordered phase, in which a microscopic symmetry of the model is spontaneously broken.
Such quantum transitions can be realized in systems of interacting fermions on $\pi$-flux or honeycomb lattices \cite{%
wang14,li15,
wang15,wang16,hesselmann16,huffman17,
he18,chen19,liu20,huffman20,schuler21,
assaad13, toldin15, otsuka16, buividovich18, lang19, liu19}, 
and may be of potential relevance for the physics of graphene-based materials~\cite{herbut06,herbut09b,liao19}.
In the Gross-Neveu transitions studied so far, all Dirac cones become simultaneously gapped out in the long-range-ordered phase. 
In the case of a continuous-symmetry breaking, this leaves behind the bosonic Goldstone modes alone as low-energy excitations. 
If only a discrete symmetry is broken, it leads to a full gap in the spectrum of the ordered phase.

\begin{figure}[b]
    \includegraphics[width=\linewidth]{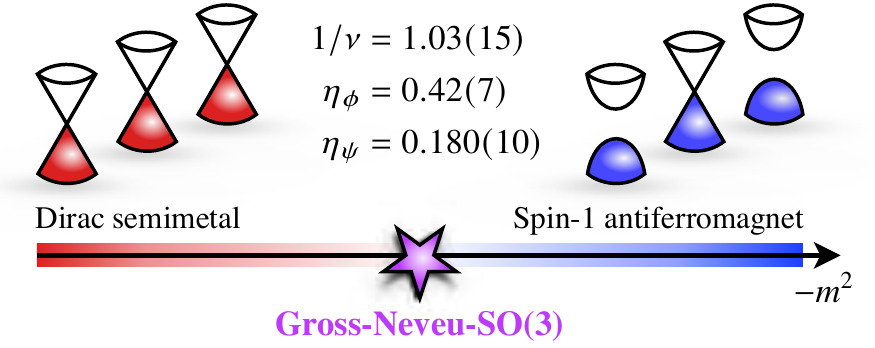}
    \caption{Quantum phase diagram of the $(2+1)$-dimensional Gross-Neveu-SO(3) model as function of tuning parameter $m^2$. The theory exhibits a quantum critical point between a Dirac semimetal and a long-range-ordered phase in which two Dirac cones acquire a mass gap, while one remains gapless, as depicted in the insets. In this work, we provide improved estimates for the universal exponents $1/\nu$, $\eta_\phi$, and $\eta_\psi$, characterizing this universality class.}
    \label{fig:phasediagram}
\end{figure}

In this work, we focus on a different family of Gross-Neveu transitions, at which the fermion spectrum is only partially gapped out. 
In particular, we study the critical behavior of the Gross-Neveu-SO(3) universality class. 
This universality class describes a transition between a symmetric Dirac semimetal phase featuring SO(3) symmetry and $N$ gapless Dirac fermions, where $N$ is an integer multiple of three, and a long-range-ordered phase, in which SO(3) is spontaneously broken and $2N/3$ Dirac cones are gapped out. Importantly, $N/3$ Dirac cones remain gapless throughout the ordered phase, as illustrated in Fig.~\ref{fig:phasediagram}.

Such a continuous quantum transition has recently been demonstrated to be realizable in frustrated spin-orbital magnets in two spatial dimensions~\cite{seifert20}. 
Here, the low-energy fermion excitations arise from fractionalization of the microscopic spin and orbital degrees of freedom. Spin-orbital models hence realize a fractionalized counterpart of the Gross-Neveu-type transitions, dubbed Gross-Neveu*.
The fractionalized Gross-Neveu* transitions differ from the ordinary Gross-Neveu transitions in the universal finite-size spectrum~\cite{schuler16, whitsitt16, seifert20}. However, in contrast to the situation in the fractionalized bosonic universality classes~\cite{isakov12, chubukov94}, at a Gross-Neveu* transition, \emph{two} independent universal exponents, such as the order-parameter anomalous dimension $\eta_\phi$ and the correlation-length exponent $\nu$, feature the same values as in the transition's ordinary counterpart.
As a consequence of the hyperscaling relations, this then implies that the exponents $\alpha$, $\beta$, $\gamma$ and $\delta$ in a fractionalized Gross-Neveu* universality class also agree with those of the corresponding ordinary Gross-Neveu universality class.
We are particularly interested in the transition between a symmetric $\mathbbm{Z}_2$ quantum spin-orbital liquid on the honeycomb lattice and a symmetry-broken phase, in which the spins order antiferromagnetically, while the orbital degrees of freedom remain disordered~\cite{seifert20}.
The $\mathbbm Z_2$ quantum spin-orbital liquid can be understood as a generalization of Kitaev's quantum spin liquid~\cite{kitaev06}, in which the number of Majorana fermions coupling to the $\mathbbm{Z}_2$ gauge field is tripled~\cite{chulliparambil20}. 
At low energy, it realizes a Dirac semimetal phase with $N=3$ two-component complex fermions and SO(3) flavor symmetry.
In the long-range-ordered phase, the SO(3) symmetry is spontaneously broken and two out of the three Dirac cones become gapped out, while the third one remains gapless. 
This partially gapped phase can be understood as a spin-1 antiferromagnet~\cite{seifert20}.

The purpose of this work is to provide refined estimates for the critical exponents characterizing the (2+1)-dimensional Gross-Neveu-SO(3) universality class.
To this end, we compare the results of three complementary advanced field-theoretical approaches.
We use a chain of computer-algebra tools developed in the context of high-energy physics~\cite{Nogueira:1991ex,Nogueira2006,Harlander1998,Seidensticker1999,vermaseren00,Kuipers2013,Ruijl2017,Czakon:2004bu,Gorishnii:1989gt,Larin1991} to determine the critical behavior within an $\epsilon$ expansion around the upper critical space-time dimension of four at three-loop order. Further,
by solving the Schwinger-Dyson equations directly at the critical point \cite{vasilev81a, vasilev81b, gracey91, graceyreview}, we compute the correlation-length exponent $\nu$ and the order-parameter anomalous dimension $\eta_\phi$ at order $\mathcal O(1/N^2)$ in the large-$N$ expansion; the fermion anomalous dimension $\eta_\psi$ is determined at order $\mathcal O(1/N^3)$ by making use of the large-$N$ conformal bootstrap technique \cite{parisi72, vasilev82, derkachov93, gracey94, gracey18}.
Finally, by employing the functional renormalization group (FRG) in the derivative-expansion scheme, we compute the critical behavior of the Gross-Neveu-SO(3) universality class at the level of the improved local potential approximation (LPA$'$).

The rest of the paper is organized as follows: In Sec.~\ref{sec:model}, we discuss the effective field theory describing the Gross-Neveu-SO(3) universality class. 
The $4-\epsilon$ and $1/N$ expansions are discussed in Secs.~\ref{sec:epsilon3} and \ref{sec:largeN}, respectively, while the FRG calculations are described in Sec.~\ref{sec:FRG}. 
{In Sec.~\ref{sec:discussion}, we present and compare the results of the three different approaches.}
The paper concludes with a short summary and outlook in Sec.~\ref{sec:conclusions}. 
Technical details are deferred to three appendices.

\section{Model}\label{sec:model}

The continuum field theory describing the Gross-Neveu-$\operatorname{SO}(3)$ universality class is given by the action $S = \int \rmd^D x \mathcal{L}$ with \cite{seifert20}\vspace{-0.2cm}
\begin{align} \label{eq:lagrangian}
    \mathcal{L} &= \bar{\psi} \gamma^\mu \partial_\mu \psi + \frac{1}{2} \phi_a \left( - \partial^2_\mu + m^2 \right) \phi_a \nonumber \\
    &\quad + \lambda (\phi_a \phi_a)^2 - g \phi_a \bar{\psi} \left(\mathds{1}_{2N/3} \otimes L_a\right)\psi
\end{align}
in $D$ Euclidean space-time dimensions. Here, we have assumed the summation convention over repeated indices $\mu = 0, \dots, D-1$ and $a=1,2,3$. 
We use conventions in which the Dirac matrices $\gamma_\mu$ form a $2N$-dimensional representation of the Clifford algebra, $\{\gamma^\mu, \gamma^\nu\} = 2 \delta^{\mu\nu} \mathds{1}_{2N}$, such that $N$ corresponds to the number of two-component fermion flavors. 
The spinor $\psi$ and its Dirac conjugate $\bar\psi \equiv \psi^\dagger \gamma^0$ have $2N$ components each.
The interaction Lagrangian comprises the $\operatorname{SO}(3)$-counterpart of the Heisenberg-Yukawa interaction~\cite{herbut09a, janssen14}, parameterized by its Yukawa coupling $g$, and a quartic boson self-interaction with coupling $\lambda$.
As in the standard Gross-Neveu-Yukawa models~\cite{hands93}, the Dirac matrices commute with the Yukawa vertex operator, $[\gamma^\mu,\mathds{1}_{2N/3} \otimes L_a] = 0$.
The $3\times 3$ matrices $L_a$ are generators of $\operatorname{SO}(3)$ in the fundamental representation, corresponding to spin 1. 
The order-parameter field $\phi_a$ is a scalar under space-time rotations, but transforms as a vector under $\operatorname{SO}(3)$. 
In $D=2$ and $D=3$ space-time dimensions, this requires that $N$ is a multiple of three, whereas in $D=4$, $N$ would need to be a multiple of six in any physical realization.
However, in what follows, it will prove to be useful to compute the critical behavior for general $2<D<4$ and arbitrary $0 \leq N \leq \infty$, allowing one to analytically continue also to noninteger values of both $D$ and $N$. 
As Aslamazov-Larkin diagrams vanish for the ungauged Gross-Neveu models~\cite{boyack19}, the critical exponents $\nu$, $\eta_\phi$, and $\eta_\psi$ do not depend on whether the theory is defined in terms of reducible or suitable copies of irreducible fermion flavors
\footnote{We note that subleading exponents, such as $\omega$, corresponding to the corrections to scaling, may depend on whether the theory is defined in terms of $N$ flavors of two-component fermions or $N/2$ flavors of four-component fermions, see Ref.~\cite{gehring15}.}.

The zero-temperature phase diagram of the Gross-Neveu-SO(3) model as a function of the tuning parameter $m^2$ can be understood on the level of mean-field theory, see Fig.~\ref{fig:phasediagram}. In this case, the fluctuations of the order parameter $\phi_a$ are neglected. Formally, this correspond to the strict limit $N \to \infty$. For $m^2 > 0$, the ground state is symmetric and the spectrum consists of $N$ gapless Dirac cones. For $m^2 < 0$, the order parameter field acquires a finite vacuum expectation value $\langle \phi_a \rangle \neq 0$ and the SO(3) flavor symmetry is spontaneously broken. However, since $L_a$ has a zero eigenvalue, only $2N/3$ of the Dirac cones acquire a mass gap, while the remaining $N/3$ Dirac cones remain gapless throughout the long-range-ordered phase.
In this work, we demonstrate that the mean-field picture remains qualitatively correct for finite values of $N$, but the corresponding critical exponents characterizing the universality class receive sizable corrections to their mean-field values.

{The field theory defined in Eq.~\eqref{eq:lagrangian} derives from a gradient expansion of a spin-orbital model on a honeycomb lattice with bond-dependent Kitaev and Heisenberg interactions at a quantum critical point between a $\mathbbm{Z}_2$ spin-orbital liquid and an antiferromagnet~\cite{seifert20}. Here, the itinerant fermionic degrees of freedom arise from fractionalization of the local moments and interact via an emergent $\mathbbm{Z}_2$ gauge field. Density matrix renormalization group calculations suggest that the gauge field excitations are gapped at the transition and thus do not contribute to the long-range behavior. Besides the order-parameter field $\phi_a$, the only low-energy degrees of freedom are therefore the gapless fermion fields $\psi$ and $\bar\psi$. The example proposed in Ref.~\cite{seifert20} corresponds to $N = 3$ two-component Dirac fermion flavors in $D=3$ space-time dimensions. However, implementations with larger values of $N$ with and without fractionalization are conceivable as well.}

\section{\texorpdfstring{$\boldsymbol{4-\epsilon}$ expansion}{4-epsilon expansion}}
\label{sec:epsilon3}

The field theory defined in Eq.~\eqref{eq:lagrangian} has an upper critical space-time dimension $D_\mathrm{up}=4$, where both, the Yukawa coupling $g$ and the quartic bosonic self-interaction $\lambda$, become simultaneously marginal. 
This allows for a controlled expansion in $D=4 - \epsilon$ dimensions. 
In this section, we report our calculation of the renormalization group functions at three-loop order. Furthermore, we extract the correlation-length exponent $\nu$, the boson anomalous dimension $\eta_\phi$, and the fermion anomalous dimension $\eta_\psi$ at order $\mathcal{O}(\epsilon^3)$.

\subsection{Method}

We define the bare Lagrangian upon replacing fields and couplings in Eq.~\eqref{eq:lagrangian} by their bare counterparts 
$\psi \mapsto \psi_0$, $\phi_a \mapsto \phi_{a,0}$, $g \mapsto g_0$ and $\lambda \mapsto \lambda_0$.
The renormalized Lagrangian reads
\begin{align}
  \mathcal{L} & = Z_\psi \bar{\psi} \gamma^\mu \partial_\mu \psi  - Z_{\phi \bar{\psi} \psi} g \mu^{\epsilon/2}\phi_a \bar{\psi}  \left(\mathds{1}_{2N/3} \otimes L_a\right)\psi  \nonumber \\
        &\quad + \frac{Z_\phi}{2} (\partial_\mu \phi_a)^2+ \frac{Z_{\phi^2}}{2} m^2 \phi_a \phi_a+ Z_{\phi^4} \lambda \mu^{\epsilon} (\phi_a \phi_a)^2\,,
\end{align}
with the renormalization constants $Z_\psi$, $Z_\phi$, $Z_{\phi\bar\psi\psi}$, $Z_{\phi^2}$, and $Z_{\phi^4}$. 
The kinetic terms in the renormalized and bare Lagrangian can be related to each other upon identifying $\psi_0 = \sqrt{Z_\psi} \psi$ and $\phi_0 = \sqrt{Z_\phi}\phi$.
The energy scale $\mu$ parametrizes the renormalization group flow. 
It is introduced upon shifting the couplings $g^2 \mapsto \mu^\epsilon g^2$ and $\lambda \mapsto \mu^\epsilon \lambda$ after the integration over $(4-\epsilon)$-dimensional spacetime.
The renormalized mass and the renormalized couplings are then related to the corresponding bare quantities as
\begin{align}
  m^2 &= m_0^2 Z_\phi Z_{\phi^2}^{-1}\,, \\
  g^2 &= g_0^2 \mu^{-\epsilon} Z_\psi^2 Z_\phi Z_{\phi\bar{\psi}\psi}^{-2}\,, \\
  \lambda &= \lambda_0 \mu^{-\epsilon} Z_\phi^2 Z_{\phi^4}^{-1}\,.
\end{align}
In the following, we compute all renormalization constants up to three-loop order. To that end, we employ dimensional regularization and the modified minimal substraction scheme ($\overline{\text{MS}}$).
This amounts to the evaluation 1,815 Feynman diagrams.
To this end, we use a sophisticated chain of computer algebra tools originally developed for loop calculations in 
high-energy physics:
First, the Feynman diagrams are generated by the program
\texttt{QGRAF}~\cite{Nogueira:1991ex,Nogueira2006}.
These are further processed by the programs \texttt{q2e} and
\texttt{exp}~\cite{Harlander1998,Seidensticker1999}, which allow one to reduce the diagrammatic expressions to 
single-scale Feynman integrals. 
Algebraic structures from the Clifford algebra and 
the SO(3) generators are contracted in \texttt{FORM}~\cite{vermaseren00,Kuipers2013,Ruijl2017}. 
Finally, the Feynman integrals are rewritten in terms of known master integrals via integration-by-parts identities~\cite{Czakon:2004bu}. Herein, the vertex functions are 
computed by setting one or two external momenta to zero and subsequently mapping to massless 
two-point functions, which are implemented in \texttt{MINCER}~\cite{Gorishnii:1989gt,Larin1991}.

\subsection{Flow equations}

The beta functions for the squared Yukawa coupling $g^2$ and the quartic scalar coupling $\lambda$ 
are defined as
\begin{align}
  \beta_{g^2} = \frac{\mathrm{d} g^2}{\mathrm{d}\ln \mu}\,, \qquad
  \beta_{\lambda} = \frac{\mathrm{d} \lambda}{\mathrm{d}\ln \mu}\,.
\end{align}
It is convenient to further rescale the couplings as $g^2/(8\pi^2) \mapsto g^2$ and $\lambda/(8\pi^2) \mapsto \lambda$,
such that the $\beta$ functions at three loop order read
\begin{align}
  \beta_{g^2} & = - \epsilon g^2 +\frac{2}{3} (N+6) g^4
  \notag\\&\quad
  - \frac{1}{2} g^2 \left[ (7 + 6 N) g^4 + 80 g^2 \lambda - 80 \lambda^2 \right]
  \notag\\&\quad
  +10 g^6 \lambda  (5 N+24)+10 g^4 \lambda ^2 (48-5 N)-440 g^2 \lambda ^3
  \notag\\&\quad 
  + 6 \zeta _3 g^8 (N+3)+\frac{1}{8} g^8 (6 N^2+37N-118)\,,\label{eq:betag2}\\
  \beta_\lambda & = -\epsilon \lambda + 44 \lambda ^2-\frac{1}{3} g^2 N (g^2-4 \lambda )
  \notag\\ 
  &\quad+ \frac{1}{3} g^2 N (5 g^4 +4 g^2 \lambda -88 \lambda ^2)-1104 \lambda ^3 
  \notag\\
  &\quad+\frac{1}{72} \Bigl\{-3 g^8 N (66 N+19) +2 g^6 \lambda  N (562 N-4761)
  \nonumber \\
  &\quad -48 g^4 \lambda ^2 N (22 N-521) +49632 g^2 \lambda ^3 N +3469248 \lambda ^4
  \nonumber \\
  &\quad -36 \zeta _3 \big[g^4 N (7 g^4+120 g^2 \lambda- 792 \lambda ^2) - 56832 \lambda ^4\big]\Bigr\}\,.\label{eq:betalam}
\end{align}
Here, $\zeta_s = \zeta(s)$ is the Riemann zeta function.
We have sorted the terms in Eqs.~\eqref{eq:betag2} and~\eqref{eq:betalam} such that the first lines show the tree level and one-loop contributions, the second lines show the two-loop contributions, and the remaining lines show the three-loop contributions.
The wave function renormalization functions $\gamma_\phi$ and $\gamma_\psi$ are defined as $\gamma_{\phi/\psi} = \mathrm{d}\ln Z_{\phi/\psi}/(\mathrm{d}\ln \mu)$. At three-loop order they read
\begin{align}
 \gamma_\phi & = \frac{2}{3} N g^2 + 40 \lambda ^2-\frac{4}{3} N g^4 
 +\frac{41 g^6 N^2}{36} 
 \notag \\&\quad 
 +\frac{g^2}{24}  N (21 g^4+400 g^2 \lambda -1200 \lambda ^2)-440 \lambda ^3\,,\\
 \gamma_\psi & = g^2 -\frac{2N+1}{4} g^4
 \notag \\&\quad
 -\frac{g^2}{48}\left[g^4 (4 N^2-84 N-9)-960 g^2 \lambda +2640 \lambda ^2\right]\,.
\end{align}
Finally, we consider the mass renormalization function as $\gamma_{\phi^2} = \mathrm{d}\ln Z_{\phi^2}/(\mathrm{d}\ln \mu)$, which at three-loop order reads 
\begin{align}
    \gamma_{\phi^2} &= -20 \lambda  -\frac{2}{3}   N g^4+\frac{40}{3} N g^2 \lambda  +240 \lambda ^2 + \frac{61  }{3}N g^6
    \notag \\&\quad
    -\frac{130}{3}N g^4 \lambda  -160 N g^2 \lambda ^2 -\frac{4}{9}N^2 g^4  (7 g^2-15 \lambda)
    \nonumber \\&\quad
    -2 \zeta _3 N g^4 (g^2+50 \lambda)-12920 \lambda ^3\,.
\end{align}
The corresponding $\beta$ function for the bosonic mass is then computed from the dimensionless mass $\tilde{m}^2 = \mu^{-2}m^2$ as
\begin{align}\label{eq:massflow}
    \beta_{\tilde{m}^2} = (2 - \gamma_\phi + \gamma_{\phi^2})\tilde{m}^2\,.
\end{align}
We note that in the limit $g^2 \rightarrow 0$,
we recover the three-loop results for the O(3)-symmetric real scalar $\phi^4$ theory~\cite{Kompaniets2017}.

\subsection{Critical exponents}

The above $\beta$ functions feature several renormalization group fixed points, i.e., coupling values $g^2_\star$ and $\lambda_\star$ at which the flow vanishes, $\beta_{g^2}(g^2_\star,\lambda_\star)=\beta_{\lambda}(g^2_\star,\lambda_\star)=0$.
At the fixed points, the system becomes scale invariant, giving rise to quantum critical behavior.
We find that the Gaussian fixed point at $(g^2_\star,\lambda_\star) = (0,0)$ and the purely bosonic Wilson-Fisher fixed point $(g^2_\star,\lambda_\star) = (0,\lambda_\star)$ are characterized by two and one relevant directions within the critical plane $\tilde m^2 = 0$, respectively. They are thus unstable and cannot be accessed in a system with a single control parameter without fine tuning.
We further find a pair of interacting fixed points at finite $g_\star^2 \neq 0$, one of which is fully infrared stable. To the leading order, the corresponding critical couplings are
\begin{align}
    \label{eq: ngfp}
    (g^2_\star,\lambda_\star) = \left( \tfrac{3 }{2 (N+6)} , \tfrac{\sqrt{N^2+120 N+36}-N+6 }{88 (N+6) } \right)\epsilon + \mathcal O(\epsilon^2)\,,
\end{align}
in agreement with the previous calculation~\cite{seifert20}.
The corresponding higher-order contributions up to $\mathcal{O}(\epsilon^3)$ are lengthy but straightforward expressions that can be obtained from Eqs.~\eqref{eq:betag2} and~\eqref{eq:betalam} analytically, and will be used in the following.

The critical behavior is determined by the renormalization group flow at and near the stable fixed point.
The anomalous dimensions are given by the wave function renormalization functions $\gamma_\psi$ 
and $\gamma_\phi$ at the fixed point,
\begin{align}
\eta_\psi = \gamma_\psi(g^2_\star,\lambda_\star)\,,\quad \eta_\phi= \gamma_\phi(g^2_\star,\lambda_\star)\,.
\end{align}
The inverse of the correlation-length exponent is extracted from the flow of the bosonic mass, which acts as tuning parameter,
\begin{align}
    \frac{1}{\nu} = \left. \frac{\mathrm{d} \beta_{\tilde{m}^2}}{\mathrm{d} \tilde{m}^2} \right|_{(g^2_\star,\lambda_\star)} = 2 - \eta_\phi + \gamma_{\phi^2}(g^2_\star,\lambda_\star)\,.
\end{align}
The full expressions for general $N$ are given in Appendix~\ref{app:epsilon}. Electronic versions of the exponents are also available as Supplemental Material for download~\cite{suppl}.

For $N=3$, which corresponds to the situation relevant for the spin-orbital models~\cite{seifert20}, the exponents read
\begin{align}
    \frac{1}{\nu} &=  2 -\tfrac{5 \sqrt{5}+9}{22} \epsilon +\tfrac{937 \sqrt{5}-3182}{31944}\epsilon ^2 
    \nonumber\\&\quad
    +\tfrac{264 \left(576665-306864 \sqrt{5}\right) \zeta _3+5132520 \sqrt{5}-113996279}{834888384 \sqrt{5}}\epsilon^3 + \mathcal O(\epsilon^4)
    \nonumber\\[8pt]
    &\approx 2 -0.917 \epsilon -0.0340 \epsilon ^2 -0.0735\epsilon ^3 + \mathcal O(\epsilon^4)\,,\label{eq: nuinv}\\[12pt]
     \eta_\phi &= \tfrac{1}{3}\epsilon+\tfrac{80 \sqrt{5}+89}{2904}\epsilon^2-\tfrac{351384 \zeta _3+66393 \sqrt{5}-357226}{6324912}\epsilon ^3 + \mathcal O(\epsilon^4)\nonumber\\[8pt]
    &\approx 0.333 \epsilon +0.0922 \epsilon ^2 -0.0338 \epsilon ^3 + \mathcal O(\epsilon^4)\,,\label{eq: etab}\\[12pt]
     \eta_\psi &= \tfrac{1}{6}\epsilon+\tfrac{105 \sqrt{5}+79}{8712}\epsilon ^2- \tfrac{234256 \zeta _3+72458 \sqrt{5}-187711}{8433216}\epsilon ^3 + \mathcal O(\epsilon^4)\nonumber\\[8pt]
    &\approx 0.167 \epsilon +0.0360 \epsilon ^2 -0.0303 \epsilon ^3 + \mathcal O(\epsilon^4)\,.\label{eq: etaf}
\end{align}
We note that the above expansions are asymptotic series with vanishing radius of convergence. It is reassuring, however, that the coefficients of the two- and three-loop corrections are still small compared to the one-loop values.
For comparison with the large-$N$ expansion, we also state the expressions that we have obtained upon further expanding the general $(4-\epsilon)$-expansion results in $1/N$. We obtain
\begin{align}
\frac{1}{\nu} &= 2-\epsilon -\left[9 \epsilon-\tfrac{39}{4} \epsilon ^2 +\tfrac{9 }{16}\epsilon ^3\right]\frac{1}{N} \nonumber\\
    &\quad +\left[459\epsilon  - \tfrac{5895}{8}\epsilon^2 + \tfrac{27}{32}(153-184\zeta _3)\epsilon^3\right]\frac{1}{N^2}\nonumber\\
    &\quad
    + \mathcal{O}(\epsilon^4,1/N^3)\,, \label{eq:epsexpansion-nu} \\
    \eta_\phi &= 
    \epsilon + \left[-6\epsilon +\tfrac{15}{4}\epsilon^2+\tfrac{21 }{16}\epsilon^3\right]\frac{1}{N}\nonumber
    \\
    &\quad +\left[36\epsilon - \tfrac{261}{8}\epsilon^2 - \tfrac{9}{32}(72\zeta _3 +95)\epsilon^3\right]\frac{1}{N^2}\nonumber\\
    &\quad+ \mathcal{O}(\epsilon^4,1/N^3)\,, \label{eq:epsexpansion-etaphi}\\
    \eta_\psi &=  \left[\tfrac{3}{2}\epsilon-\tfrac{9}{8}\epsilon^2-\tfrac{9}{32}\epsilon^3\right]\frac{1}{N} 
        +\left[-9\epsilon+\tfrac{369}{16}\epsilon^2-\tfrac{513}{64}\epsilon^3\right]\frac{1}{N^2} \nonumber\\
        &\quad + \left[54\epsilon - \tfrac{4023}{16}\epsilon^2 + \tfrac{243}{32}(33-4\zeta _3)\epsilon ^3\right]\frac{1}{N^3}\nonumber\\ 
        &\quad + \mathcal{O}(\epsilon^4,1/N^4)\,. \label{eq:epsexpansion-etapsi}
\end{align}
For any fixed $N$, we extract estimates for the physical dimension $\epsilon = 1$ by employing standard Pad\'e approximants 
\begin{align}
    [m/n] = \frac{a_0 + a_1 \epsilon + \dots + a_m \epsilon^m}{1+ b_1 \epsilon + \dots + b_n \epsilon^n}\,,
\end{align}
with $m, n \in \{0, 1, 2, 3\}$ and $m+n=3$. The coefficients $a_0, \dots, a_m$ and $b_1, \dots, b_n$ are obtained from matching the Taylor series of $[m/n]$ order by order with the $\epsilon$ expansions.
The discussion of the resulting estimates for $1/\nu$, $\eta_\phi$, and $\eta_\psi$ for different values of $N$ is deferred to Sec.~\ref{sec:discussion}.

\section{\texorpdfstring{$\boldsymbol{1/N}$ expansion}{1/N expansion}}
\label{sec:largeN}

In the limit of a large number of fermion flavors $N \to \infty$, the
fluctuations of the order-parameter field $\phi_a$ freeze out, which allows us
to compute the critical exponents in arbitrary $2 < D < 4$ in a
systematic expansion in powers of $1/N$; this is the topic of the present section.

\subsection{Method}

To achieve this, we have applied the large-$N$ critical point method developed
in Refs.~\cite{vasilev81a,vasilev81b,vasilev82} for the scalar O($N$) model and
later extended to the Gross-Neveu universality class in
Refs.~\cite{gracey91,derkachov93,vasilev93a,vasilev93b,gracey93,gracey94}.
As the latter formalism has already been applied to variations of the
Gross-Neveu model, we will highlight only the key differences here. Indeed given the strong overlap
with the Gross-Neveu-SU(2) (= chiral Heisenberg) model that the present SO(3)
study is similar to, we refer the reader to Ref.~\cite{gracey18} for the finer
details of the technique.

One of the first steps is to recognize that the Lagrangian which serves as the
basis for the method of Refs.~\cite{vasilev81a,vasilev81b,vasilev82} is that of the
universal theory that resides at the stable fixed point in all dimensions $2<D<4$. 
It is a simpler version of Eq.~(\ref{eq:lagrangian}) in that
only the fermion kinetic term and the three-point vertex are the essential ones
needed to define the canonical dimensions of the fields at the fixed point,
together with a quadratic term in the boson field. Specifically,
\begin{equation}
\mathcal L_{\mathrm{univ}} = \bar{\psi} \slashed{\partial} \psi -
{\phi}_a \bar{\psi} \left( \mathds{1}_{2N/3} \otimes L_a \right)
\psi + \frac{1}{2} {\phi}_a {\phi}_a\,,
\label{laguniv}
\end{equation}
where $\slashed{\partial} \equiv \gamma^\mu \partial_\mu$ with $\gamma^\mu$ again being $(2N)\times(2N)$ Dirac matrices, such that the spinors $\psi$ and $\bar\psi$ have $2N$ components, as in the original Lagrangian [Eq.~\eqref{eq:lagrangian}].
The scalar ${\phi}_a$ has been rescaled since at
criticality the perturbative coupling constant is fixed and does not run. The
quartic interaction present in Eq.~(\ref{eq:lagrangian}) is required in four dimensions to
ensure renormalizability. Its contribution in $\mathcal L_{\mathrm{univ}}$ is automatically
accounted for through closed fermion loop diagrams with four external boson
fields~\cite{hasenfratz2}. The other main aspect of the setup concerns the
algebra of the SO(3) generators $L_a$, which satisfy the relation
\begin{equation}
(L_{a})_{ij} (L_{a})_{kl} = \delta_{il} \delta_{jk} - \delta_{ik} \delta_{jl}\,.
\end{equation}
We have used this in determining the group-theory factors associated with the
Feynman diagrams that contribute to the large-$N$ formalism.

In general, the method of Refs.~\cite{vasilev81a,vasilev81b,vasilev82} entails
analyzing the behavior of various Schwinger-Dyson equations in the approach to
criticality. At the stable fixed point, the propagators of the fields
have a simple scaling behavior where the exponent of the propagator corresponds to the full
scaling dimension. 
Specifically, in coordinate space the propagators take the asymptotic forms
\begin{align}
\psi(x) &\sim \frac{A\slashed{x}}{(x^2)^\alpha} \left[ 1 + A^\prime(x^2)^\lambda
\right], 
\label{propcrit-psi}\\
\phi(x) &\sim \frac{B}{(x^2)^\beta} \left[ 1 + B^\prime(x^2)^\lambda
\right],
\label{propcrit-phi}
\end{align}
where we have used the name of the field as a shorthand for the propagator at criticality, with the scaling exponents
\begin{align}
\alpha & = \tfrac12 (D + \eta_\psi)\,, &
\beta & = 1 - \eta_\psi - \chi\,.
\label{critdim}
\end{align}
Here, $\eta_\psi$ is the fermion anomalous dimension, which has been computed to three loops at
criticality in the previous section. The anomalous dimension of the boson-fermion
vertex is denoted by $\chi$ so that
\begin{equation}
\eta_\phi = 4 - D - 2 \eta_\psi - 2 \chi\,.
\label{etabdef}
\end{equation}
In addition to these leading exponents, each propagator includes a
correction term involving the exponent $\lambda$. At criticality, this exponent
corresponds to the correlation-length exponent as $1/\nu = 2\lambda$.
The canonical dimension of $\lambda$ is
$(D-2)/2$. The quantities $A$, $B$, as well as $A^\prime$ and $B^\prime$ are
$x$-independent amplitudes. The first two always appear in the combination
$A^2 B$, but this plays an intermediate role in deriving exponents. The first
terms of the respective equations in Fig.~\ref{sde2pt} represent the asymptotic
scaling forms of the two-point functions and have been given in 
Ref.~\cite{gracey91}. They are derived from Eqs.~\eqref{propcrit-psi} and 
\eqref{propcrit-phi} and have a similar
scaling form to these, although $A$ and $B$ occur in the denominator.

\subsubsection{Skeleton Schwinger-Dyson equations}

{\begin{figure}[tb]
\begin{center}
\includegraphics[width=8.5cm,height=2.5cm]{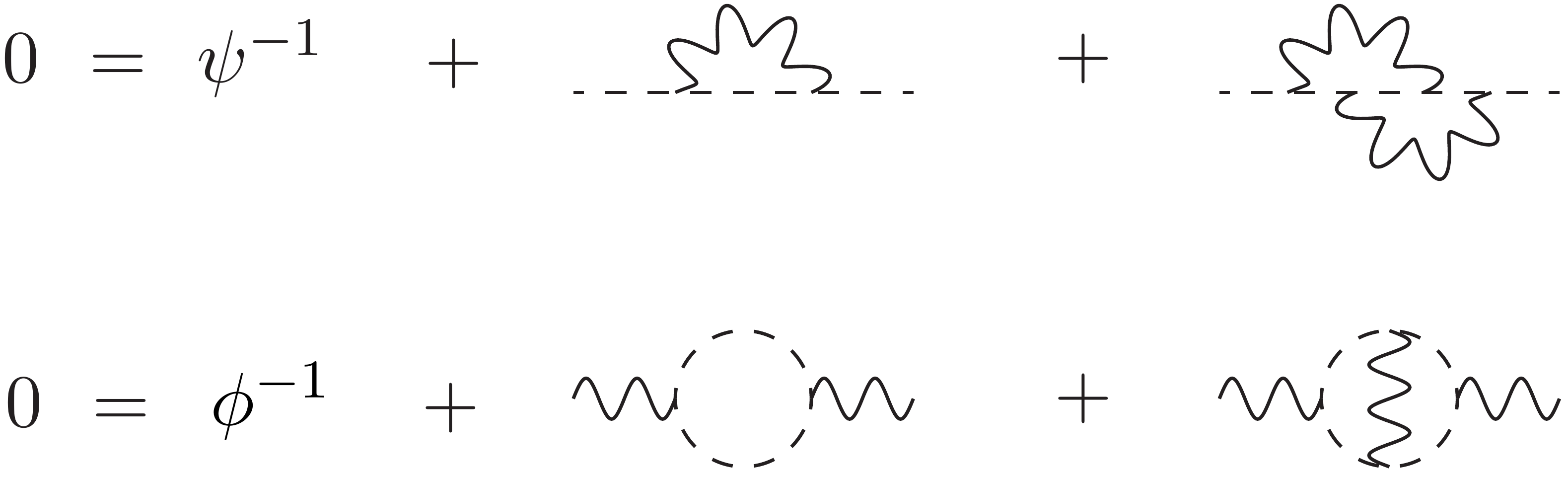}
\end{center}
\caption{Skeleton Schwinger-Dyson two-point functions used to determine
$\eta_\psi$ at $\mathcal O(1/N^2)$. Dashed inner lines correspond to critical fermion propagators [Eq.~\eqref{propcrit-psi}] and wiggly inner lines correspond to critical boson propagators [Eq.~\eqref{propcrit-phi}].
} 
\label{sde2pt}
\end{figure}}

\begin{figure}[tb]
\begin{center}
\includegraphics[width=1.5cm,height=1.5cm]{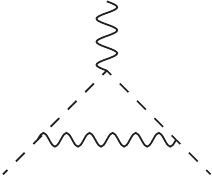}
\end{center}
\caption{Leading-order skeleton Schwinger-Dyson three-point function used to
determine $\chi$ at $\mathcal O(1/N)$.}
\label{sde3pt}
\end{figure}

\begin{figure}[tb]
\begin{center}
\includegraphics[width=8.5cm,height=5.5cm]{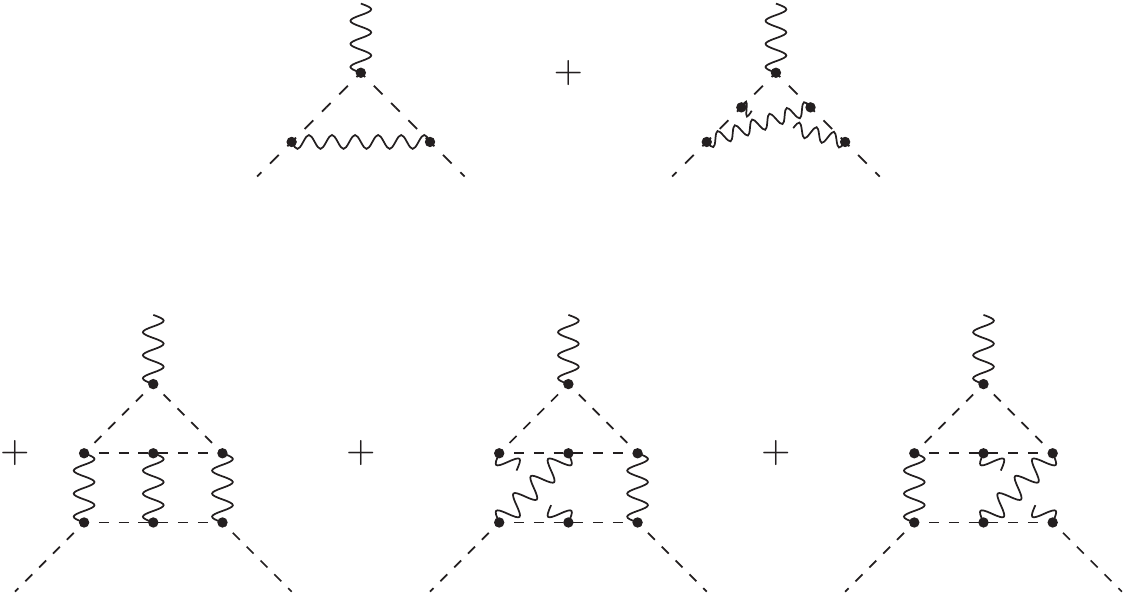}
\end{center}
\caption{Diagrams contributing to large-$N$ conformal bootstrap formalism to
deduce $\eta_\psi$ at $\mathcal O(1/N^3)$. Black dots refer to Polyakov conformal triangles, see Ref.~\cite{gracey18} for details.}
\label{3ptcf}
\end{figure}

To determine the anomalous dimensions of the two fields, one focuses on the
two-point Schwinger-Dyson equations shown in Fig.~\ref{sde2pt}, as well the three-point
vertex function, for which the first correction is depicted in Fig.~\ref{sde3pt}. 
For both the two- and three-point functions the contributing diagrams
are computed with the asymptotic propagators, Eqs.~\eqref{propcrit-psi} and \eqref{propcrit-phi}. Since the power of
the leading term of each propagator includes the nonzero anomalous dimensions
of Eq.~(\ref{critdim}), there are no self-energy corrections on the
contributing diagrams in order to avoid double counting. By evaluating the diagrams
and solving the equations of Fig.~\ref{sde2pt} self-consistently (eliminating
the product $A^2 B$ in the process), one obtains an expression for $\eta_\psi$ at $\mathcal O(1/N^2)$. The
value of $\chi$ at $\mathcal O(1/N)$ is required for this to ensure that no $\ln(x^2)$
terms remain after renormalization. This value for $\chi$ is deduced from the
scaling behavior of the diagram of Fig.~\ref{sde3pt}. Moreover, this produces
$\eta_\phi$ at $\mathcal O(1/N)$ as a corollary from Eq.~(\ref{etabdef}). For the next order of
$\chi$, one extends the critical-point evaluation of the higher-order diagrams to
the three-point function, which are given by the decorations of the leading-order diagram of Fig.~\ref{sde3pt}
with vertex corrections, as well as the non-planar and three-loop
diagrams shown in Fig.~\ref{3ptcf}. This produces $\chi$ and hence $\eta_\phi$ at
$\mathcal O(1/N^2)$.

Once we have established the anomalous dimensions of the fields at $\mathcal O(1/N^2)$,
the correction to scaling terms in Eqs.~(\ref{propcrit-psi}) and (\ref{propcrit-phi}) can be included in order to
determine $1/\nu$ via the determination of $\lambda$. Since the correction
terms involve $(x^2)^\lambda$, the two-point Schwinger-Dyson consistency
equation contains terms of different dimensions. These split into terms which
are independent of the correction to scaling amplitudes, $A^\prime$ and
$B^\prime$, and those that are not. It is the latter ones that determine
$\lambda$ to $\mathcal O(1/N^2)$ \cite{vasilev81b}, since a consistency equation can be
formed from the $2 \times 2$ matrix defined by the coefficients of
$A^\prime$ and $B^\prime$ in each equation of Fig.~\ref{sde2pt}. Finding the
solution to the equation formed by setting the determinant of this matrix to
zero defines the consistency equation. For the Gross-Neveu universality classes
there is a known complication in that while all the propagators of the diagrams
of Fig.~\ref{sde2pt} include the correction terms, extra diagrams are needed due
to the same reordering that arises in the original Gross-Neveu-$\mathbbm Z_2$ (= chiral Ising)
model~\cite{gracey91,derkachov93,vasilev93b,gracey93}. This
necessitates the inclusion of the higher-order Feynman diagram as given in Fig.~$4$
of Ref.~\cite{gracey18}, but with the appropriate group factor for the present model.

\subsubsection{Large-$N$ conformal bootstrap technique}

Finally, we have been able to apply what is termed the large-$N$ conformal
bootstrap technique to compute the $\mathcal O(1/N^3)$ term of $\eta_\psi$. This method was
originally developed for the O($N$) scalar model in Ref.~\cite{vasilev82} using the
early work of Refs.~\cite{polyakov70,parisi72,deramo72}. It was subsequently extended
to the Gross-Neveu-$\mathbbm Z_2$ universality class in 
Refs.~\cite{derkachov93,vasilev93a,gracey94} and more recently to the Gross-Neveu-SU(2) (= chiral
Heisenberg) model in Ref.~\cite{gracey18} and the Gross-Neveu-U(1) (= chiral XY) model in Ref.~\cite{gracey21}. We refer readers to that later article for more details of the large-$N$ conformal
bootstrap technique for the present context. However, it is
worth noting some of the key aspects of the approach. Rather than focusing on the
skeleton Schwinger-Dyson two-point functions, the underlying self-consistency
equations that ultimately produce $\eta_\psi$ at $\mathcal O(1/N^3)$ are derived from the vertex
functions. By contrast to the two-point function approach, one is in effect
performing perturbation theory in the vertex anomalous dimension $\chi$. The
relevant diagrams are given in Fig.~\ref{3ptcf}. Again, while there is no dressing
on the propagators, there are no vertex corrections unlike the diagrams in Fig.~\ref{sde2pt}. 
Instead, the contributions that underlie the vertex structure are
subsumed into the black dots, which denote Polyakov conformal triangles. These are
designed in such a way that the sum of the critical exponents of the
propagators connected to the vertex is $(D + 1)$. This value means that all
the scalar-fermion vertices are unique in the sense of conformal 
integration~\cite{gracey91,derkachov93,vasilev93b,gracey93}. It is hence possible to evaluate
all the diagrams to the necessary order to determine $\eta_\psi$ at $\mathcal O(1/N^3)$.

\subsection{Critical exponents}

Having summarized the large-$N$ critical point formalism, we are now in a
position to discuss the results. Expressions in general space-time dimensions $2<D<4$ for all the
exponents we have determined are presented in Appendix~\ref{app:large-N} and electronically as Supplemental Material~\cite{suppl}.
However, the
$\epsilon$ expansion of the large-$N$ expressions must agree with the explicit 
three-loop exponents derived from the renormalization group functions at the
stable fixed point. Therefore, if we expand each of $\eta_\psi$, $\eta_\phi$,
and $1/\nu$ around $D=4-\epsilon$, we find
\begin{widetext}
\begin{align}
 \frac{1}{\nu} &= 2 - \epsilon + \left[ - 9 \epsilon
+  \tfrac{39}{4} \epsilon^2 -  \tfrac{9}{16} \epsilon^3
-  \tfrac{9}{64} ( 1 + 16 \zeta_3 ) \epsilon^4
+  \tfrac{3}{256} ( 208 \zeta_3 - 144 \zeta_4 - 3 ) \epsilon^5
\right]  \frac{1}{N} 
\nonumber \\ &\quad 
+ \left[ 459 \epsilon -  \tfrac{5895}{8} \epsilon^2
+  \tfrac{27}{32} ( 153 - 184 \zeta_3 ) \epsilon^3
+  \tfrac{27}{64} ( 320 \zeta_5 - 276 \zeta_4 + 1376 \zeta_3 + 203 ) \epsilon^4
\right.\nonumber\\&\quad\quad\left.
+  \tfrac{9}{1792} ( 4795 - 90496 \zeta_5 + 6720 \zeta_3^2 - 123984 \zeta_3
+ 33600 \zeta_6 + 86688 \zeta_4 ) \epsilon^5
\right]  \frac{1}{N^2} 
+ \mathcal O ( \epsilon^6, {1}/{N^3})\,,
\displaybreak[0] \\
\eta_\phi &= \epsilon + \left[ - 6 \epsilon +  \tfrac{15}{4} \epsilon^2
+  \tfrac{21}{16} \epsilon^3 +  \tfrac{3}{64} ( 11 - 32 \zeta_3 ) \epsilon^4
+  \tfrac{3}{256} ( 80 \zeta_3 - 96 \zeta_4 + 19 ) \epsilon^5
\right]  \frac{1}{N} 
\nonumber \\&\quad 
+ \left[ 36 \epsilon -  \tfrac{261}{8} \epsilon^2
-  \tfrac{9}{32} ( 72 \zeta_3 + 95 ) \epsilon^3
+  \tfrac{9}{64} ( 472 \zeta_3 - 108 \zeta_4 + 45 ) \epsilon^4
+  \tfrac{9}{256} ( 97 - 288 \zeta_5 + 1416 \zeta_4 - 1248 \zeta_3 ) \epsilon^5
\right]  \frac{1}{N^2}
\nonumber \\ &\quad
+ \mathcal O ( \epsilon^6, {1}/{N^3})\,,
\displaybreak[0]\\
\eta_\psi &= \left[  \tfrac{3}{2} \epsilon -  \tfrac{9}{8} \epsilon^2
-  \tfrac{9}{32} \epsilon^3 +  \tfrac{3}{128} ( 16 \zeta_3 - 3 ) \epsilon^4
+  \tfrac{9}{512} ( 16 \zeta_4 - 16 \zeta_3 - 1 ) \epsilon^5
\right]  \frac{1}{N} \nonumber \\
&\quad + \left[ - 9 \epsilon +  \tfrac{369}{16} \epsilon^2
-  \tfrac{513}{64} \epsilon^3 -  \tfrac{9}{128} ( 128 \zeta_3 + 69 ) \epsilon^4
+  \tfrac{9}{512} ( 1008 \zeta_3 - 384 \zeta_4 - 89 ) \epsilon^5
\right]  \frac{1}{N^2} \nonumber \\
&\quad + \left[ 54 \epsilon -  \tfrac{4023}{16} \epsilon^2
+  \tfrac{243}{32} ( 33 - 4 \zeta_3 ) \epsilon^3
+  \tfrac{27}{256} ( 2184 \zeta_3 - 216 \zeta_4 + 493 ) \epsilon^4
+  \tfrac{27}{1024} ( 6552 \zeta_4 - 576 \zeta_5 - 20024 \zeta_3 - 2375 )
\epsilon^5 \right] \frac{1}{N^3}
\nonumber\\&\quad
+ \mathcal O (\epsilon^6, {1}/{N^4})\,.
\end{align}
\end{widetext}
All terms to $\mathcal O(\epsilon^3)$ agree exactly with Eqs.~(\ref{eq:epsexpansion-nu})--\eqref{eq:epsexpansion-etapsi}, which
is a highly non-trivial check on our $D$-dimensional expressions. In the above equations, we have
included additional terms to $\mathcal O(\epsilon^5)$ to provide
checks for future higher-loop computations.

With this check of the $D$-dimensional exponents satisfied, we can now deduce
their values in the $1/N$ expansion in fixed $D=2+1$ space-time dimensions. We find
\begin{align}
\frac{1}{\nu} &= 1  - \frac{16}{\pi^2 N}  + 
\frac{324\pi^2 + 2624}{3\pi^4 N^2}
+  \mathcal O ({1}/{N^3})
\nonumber \\
& \approx
1  -  \frac{1.62114}{N}  +  \frac{19.92200}{N^2}  + 
\mathcal O ({1}/{N^3})\,,
\displaybreak[0]\\
\eta_\phi &= 1  - \frac{20}{\pi^2 N}  +  \frac{2(81\pi^2 - 1028)}{3\pi^4 N^2}  
+ \mathcal O ({1}/{N^3})
\nonumber \\
& \approx
1  -  \frac{2.02642}{N}  +  \frac{1.56428}{N^2}  + 
\mathcal O ({1}/{N^3})\,,
\displaybreak[0]\\
\eta_\psi &= \frac{4}{\pi^2 N}  +  \frac{304}{3\pi^4 N^2} 
\nonumber \\ &\quad
+ \frac{972 \pi^2 \ln(2) + 255 \pi^2 - 10206 \zeta_3 - 3796}{9\pi^6 N^3} 
+ \mathcal O ({1}/{N^4})
\nonumber \\
& \approx
\frac{0.40528}{N}  +  \frac{1.04029}{N^2}  - 
\frac{0.79721}{N^3}  +  \mathcal O ({1}/{N^4})\,.
\end{align}
In effect, three terms in the expansion of each exponent are
available, but involve different powers of $1/N$. We note that the leading two terms of $1/\nu$ and the leading terms of
$\eta_\phi$ and $\eta_\psi$ are the same as those of the Gross-Neveu-SU(2) model~\cite{gracey18}.
However, the $\mathcal O(1/N^2)$ term of $1/\nu$ is nearly
twice that of its SU(2) counterpart and the coefficients of the subsequent terms
of $\eta_\phi$ and $\eta_\psi$ are also significantly larger here, with the exception of the $\mathcal O(1/N^2)$ term in $\eta_\phi$.

For extrapolating the large-$N$ series to finite $N$, we again use Pad\'e approximants
\begin{align}
    [m/n] = \frac{a_0 + a_1 N^{-1} + \dots + a_m N^{-m}}{ 1 + b_1 N^{-1} + \dots + b_n N^{-n}}\,,
\end{align}
where now $m,n \in \{0,1,2\}$ ($m,n \in \{0,1,2,3\}$) and $m+n=2$ ($m+n=3$) for $1/\nu$ and $\eta_\phi$ ($\eta_\psi$). The numerical estimates for different values of $N$ are discussed in Sec.~\ref{sec:discussion}.

\section{Functional renormalization group}
\label{sec:FRG}

Finally, as the physical case of interest $D=2+1$ and $N=3$ lies outside the regimes in which the $\epsilon$ and $1/N$ expansions are strictly controlled, we also employ the FRG as a complementary approach to estimate the critical exponents.

\subsection{Method}\label{sec:FRGmethod}

The FRG is a method to compute the quantum effective action $\Gamma[\Phi]$, which is the generating functional of one-particle irreducible (1PI) Green's functions~\cite{bergesreview}. Here, $\Phi$ corresponds to a collective field variable, which comprises all individual fields contained in the theory.
In the Gross-Neveu-$\operatorname{SO}(3)$ case, we have $\Phi = (\phi_a,\psi,\bar{\psi})$.
The effective action contains all quantum fluctuations, in the sense that
\begin{align}
    \Gamma[\Phi] = -\ln \int_{\text{1PI}}\mathcal{D}\Phi^\prime \rme^{-S[\Phi + \Phi^\prime]},
    \label{eq:Gamma1PI}
\end{align}
where $S$ refers to the microscopic action and the subscript reminds that only 1PI diagrams are allowed to contribute to the path integral. The key idea of the renormalization group approach is to perform this integration step by step. To this end, one extends the action $S$ by a scale-dependent regulator term,
\begin{align}
    S \mapsto S_k = S + \int_q \Phi'(-q) {R}_k(q) \Phi'(q)
    \label{eq:Gammak1PI}
\end{align}
which leads to the corresponding effective \emph{average} action $\Gamma_k[\Phi]$.
The raison d'\^etre of the regulator ${R}_k(q)$ is to suppress ``slow'' fluctuation modes $\Phi'(q)$ with momenta $q \lesssim k$ in the path integral; as such, it needs to satisfy ${R}_k(q \ll k) = C_k >0$ for $k>0$, with $C_{k\to \infty} \to \infty$.
For $k \to 0$, we demand that all modes should be integrated out and thus ${R}_{k \to 0}(q) = 0$ for all momenta $q$. 
The average action $\Gamma_k$ then interpolates between the microscopic action $S$ at the ultraviolet cutoff $k \to \Lambda \to \infty$ and the full quantum effective action $\Gamma$ in the infrared limit $k \to 0$.
The advantage of the scale-dependent formulation is that the 1PI path-integral prescription for $\Gamma_k$ can be traded for an evolution equation in functional space, to wit
\begin{align}
   \partial_t \Gamma_k = \frac12 \operatorname{Str}\frac{\partial_t {R}_k}{\Gamma_k^{(2)} + {R}_k},
   \label{eq:Wetterich}
\end{align}
which is known as the Wetterich equation~\cite{wetterich93}. Here, we have introduced the scale derivative $\partial_t \equiv k\partial_k$, $\Gamma_k^{(2)}$ is the Hessian
\begin{align}
    \Gamma_k^{(2)} = \frac{\overrightarrow{\delta}}{\delta \Phi^\top} \Gamma_k \frac{\overleftarrow{\delta}}{\delta \Phi},
\end{align}
and the supertrace operator $\operatorname{STr}$ extends the usual trace by accounting for Fermi-Dirac statistics thus:
\begin{align}
    \operatorname{STr}
    \begin{pmatrix}
    B & * & * \\
    * & F_1 & * \\
    * & * & F_2
    \end{pmatrix} = \operatorname{Tr}B - \operatorname{Tr}
    \begin{pmatrix}
    F_1 & * \\
    * & F_2
    \end{pmatrix}.
\end{align}
See Refs.~\cite{bergesreview, kopietzbook, giesreview, wipfreview} for introductory expositions on the method, and Refs.~\cite{metznerreview, braunreview, dupuisreview} for reviews on applications to interacting many-body systems.
The Wetterich equation itself is exact, but generically not exactly soluble. 

In the absence of a small control parameter for the physical case of $N=3$ and $D=3$, here we pursue an ansatz in the spirit of a derivative expansion of the effective average action,
\begin{align}
    \Gamma_k &= \int \rmd^D x \left[Z_{\psi,k} \bar{\psi} \gamma^\mu \partial_\mu \psi + \frac12 Z_{\phi,k} (\partial_\mu \phi_a)^2 \right. \nonumber\\
    &\phantom{{}=} \left. {}\vphantom{\frac12}- g_k \phi_a\bar{\psi} (\idCliff \otimes L_a) \psi + U_k(\varrho)\right],
    \label{eq:LPAprimeansatz}
\end{align}
where we have introduced the $\operatorname{SO}(3)$-invariant $\varrho = \frac12\phi_a\phi_a$. General field-dependence of renormalization group functions is allowed only in the effective average bosonic potential $U_k$, which is assumed to carry no explicit momentum dependence. Pure fermionic interactions, such as four-fermion terms, that may be generated in the nonperturbative regime, are neglected. The next-to-leading order contributions come from the kinetic terms, whose scale-dependences are approximated by field-independent renormalization constants $Z_{\Phi,k}$; all higher-order terms in the gradient expansion are neglected. This truncation of the effective average action is commonly referred to as ``improved local potential approximation'' (LPA$'$). 
It has been proven to yield reliable results in a number of similar Gross-Neveu-Yukawa-type models~\cite{rosa01, hoefling02, rechenberger10, braun11, scherer13, janssen14, classen16, classen17, janssen17, torres18, torres20}.
Extensions of this approximation for the present class of models have been discussed in Refs.~\cite{janssen12, vacca15, knorr16, knorr18}.
A final approximation entails choosing a suitable ansatz for the effective average potential.
Here, we employ two different expansion techniques; we have verified that our numerical results from the two approaches converge to the same values within the error bars.

\subsubsection{Taylor expansion of effective potential}

A simple ansatz is a truncated Taylor expansion
\begin{align} \label{eq:polynomial-truncation}
    U_k(\varrho) = \sum_{i=1}^{n/2} \frac{1}{i!}\lambda_{i,k}\, \varrho^i,
\end{align}
where we have assumed that the fixed point is located in the symmetric regime, such that the minimum of the potential is at $\varrho = 0$. If this assumption is violated at the fixed point, i.e., $U'(0) < 0$, an alternative expansion
\begin{align}
    U_k(\varrho) = \sum_{i=2}^{n/2} \frac{1}{i!}\hat{\lambda}_{i,k}\, (\varrho - \varrho_{0,k})^i
\end{align}
is more expedient; this is called the spontaneously symmetry broken (SSB) regime. In the above, $\varrho_{0,k}$ is the (scale-dependent) location of the minimum of $U_k(\varrho)$. It is related to the vacuum expectation value (VEV) of the order parameter by $\rho_{0,0} = \frac12\langle \phi_a \rangle^2$. Note that the linear term in the Taylor expansion is absent, since $\partial U(\varrho)/\partial \phi_a = \phi_a U'(\varrho)$, and hence $U'(\varrho_{0}) = 0$ if $\varrho_{0} \neq 0$ is a local minimum.

For practical computations, the ansatz~\eqref{eq:polynomial-truncation} is truncated at some finite order $n \in 2\mathds{N}$. This defines the so-called LPA$n'$. The validity of this polynomial truncation can be checked {\it a posteriori} by verifying convergence of the results upon increasing $n$. The expansion of the effective potential introduces a plethora of coupling constants, of which $\lambda_1 = m^2 > 0$ is proportional to the squared boson mass and $\lambda_2 = 4!\lambda$ is the quartic boson self-coupling. Inclusion of the higher-order couplings $\lambda_{i > 2}$ is a minimal way to incorporate nonperturbative corrections in space-time dimensions $D<4$, in addition to the effects from the nonperturbative propagator, cf.~Eq.~\eqref{eq:Wetterich}.

The flow of the bosonic self-couplings are determined from the flow of $U_k(\varrho)$ by differentiating successively with respect to $\varrho$. In the symmetric regime, this is straightforward to implement:
\begin{align}
    \partial_t \lambda_i = \left[(\partial_\varrho)^i\partial_t U_k(\varrho)\right]_{\varrho \to 0} \quad (i \in \mathds{N}_{\geqslant 1}).
\end{align}
The corresponding system of equations in the SSB regime is given by
\begin{align}
    \partial_t \hat{\lambda}_i = \left[(\partial_\varrho)^i\partial_t U(\varrho)\right]_{\varrho \to \varrho_0} + \hat{\lambda}_{i+1}\partial_t \varrho_0 \quad (i \in \mathds{N}_{\geqslant 2}),
\end{align}
and has to be supplemented by a flow equation for the VEV:
\begin{align}
    \partial_t \varrho_0 = -\frac{1}{\hat{\lambda}_2} \left[\partial_\varrho \partial_t U(\varrho)\right]_{\varrho \to \varrho_0}.
\end{align}
The latter follows from $U'(\varrho_0) = 0$ in the SSB regime \cite{rechenberger10}.

\subsubsection{Pseudospectral decomposition of effective potential}\label{sec:pseudospec}

In the context of the present work, we aim at systematically comparing the results from different quantum-field-theoretical methods between two and four dimensions. 
In particular, towards two dimensions, we have to be careful about a possible breakdown of the convergence of a local expansion in the effective potential.
This is related to the canonical dimensionality of the operators or couplings in the local expansion, i.e., the terms $\propto \lambda_i \varrho^i$.
More specifically, the canonical dimension $[\cdot]$ of the bosonic field $\phi$ is given by $[\phi]=(D-2)/2$, i.e., the dimension of the operator $\varrho^i$ is $(D-2)i$.
Therefore, the corresponding coupling $\lambda_i$ scales as $[\lambda_i]=D-(D-2)i$.
Lowering the dimension towards $D=2$ means that more and more couplings with higher $i$ become canonically relevant until they all have the same canonical dimension of two in $D=2$.
Depending on the model and the specific fixed point, this behavior can severely limit the reliability of a finite-order local expansion in the bosonic operators.

In lieu of a local Taylor expansion for the effective potential, non-local expansion schemes can be advantageous in terms of tractability, accuracy, and fast convergence.
An approximation scheme that has been explored in the context of FRG fixed-point and flow equations is based on pseudospectral methods~\cite{boyd01}.
Importantly, these methods facilitate, e.g., an efficient and high-precision resolution of global aspects of the effective potential including the correct description of a model's asymptotic behavior~\cite{litim03,fischer04,borchardt15,borchardt16a,borchardt16b,borchardt16c,knorr16,knorr18}.

In the present case of the fixed point equation for the effective potential, we need to find an approximate solution to an ordinary differential equation in one variable defined on the domain $\mathbb{R}^+ \coloneq [0,\infty)$.
To that end, we can expand the effective potential $U(\varrho)$ into a series of Chebyshev polynomials, where the domain of $U(\varrho)$ is decomposed into two subdomains, i.e., $[0,\varrho_{\rm m}]$ and $[\varrho_{\rm m},\infty)$.
The expansion then reads as
\begin{align}\label{eq:cheby}
	U(\varrho)\approx 
	\begin{cases} 
	\sum_{i=0}^{n_T}t_iT_i\left(\frac{2\varrho}{\varrho_{\rm m}}-1\right)\,,\quad \varrho \leq \varrho_{\rm m}\,,\\[5pt]
	U_{\infty}(\varrho)\sum_{i=0}^{n_R}r_iR_i(\varrho-\varrho_{\rm m})\,,\quad \varrho \geq \varrho_{\rm m}\,.
	\end{cases}
\end{align}
Here, the $T_i(x)$ are the Chebyshev polynomials of the first kind, and the $R_i(x)=T_i\big(\frac{x-L}{x+L}\big)$ are rational Chebyshev polynomials with a free parameter $L$ which parameterizes the compactification in the argument~$x$.
Further, $U_\infty(\varrho)$ is the leading asymptotic behavior of the effective potential for large field arguments, i.e., $\varrho\to\infty$, which we obtain from the dimensional scaling terms in the flow equation.
The matching point $\varrho_{\rm m}$ separates the subdomains and is another free parameter that has to be chosen large enough such that the minimum of the effective potential appears for $\varrho=\varrho_0 < \varrho_{\rm m}$. 
We can use $L$ and $\varrho_{\rm m}$ to optimize numerical convergence. 
The values of the effective potential and its derivatives for all field arguments $\varrho$ are straightforwardly obtained by employing efficient recursive algorithms~\cite{boyd01}.
In fact, we only need a relatively small number of expansion coefficients $t_i$ and $r_i$ due to fast convergence of the series.

For the determination of the coefficients $t_i$ and $r_i$ in the Chebyshev expansion, we use the collocation method, i.e., we insert the ansatz in Eq.~\eqref{eq:cheby} into the flow Eq.~\eqref{eq:Wetterich} and evaluate it on a given set of collocation points.
The collocation points are chosen to be the nodes of the highest Chebyshev polynomials in the respective domain, and we add the origin $\varrho=0$.
Finally, to accomplish smoothness, we implement matching conditions for the values of the effective potential and its derivatives at $\varrho_{\rm m}$.
The resulting set of algebraic equations is then solved with the Newton-Raphson method.
In practice, we actually expand the derivative of the dimensionless effective potential $u'(\varrho)$ along these lines, and we optimize $L$ and $\varrho_{\rm m}$ as well as the number of collocation points until we reach convergence in our numerical results. For the present model, we observe numerical convergence of the first four significant digits already starting at $n_T=n_R=9$ and, as a sanity check, we have increased the number of collocation points up to 18 in each subdomain for selected cases.

The anomalous dimensions of the quantum critical point are then obtained directly from the fixed-point solution of $u'(\varrho)$ using the FRG flow equations specified in the next section.
To obtain the inverse correlation-length exponent, we use the pseudospectral expansion from the first subdomain, i.e., $\varrho < \varrho_{\rm m}$, rewriting it as a local expansion around its minimum.
With the latter expansion, we then calculate the stability matrix and extract the eigenvalues at the fixed-point potential. The largest positive eigenvalue is the inverse correlation-length exponent.

\subsection{Flow equations}

For convenience, we introduce dimensionless versions of renormalized couplings and the effective potential, to wit:
\begin{align}
    \tilde{g}^2 = Z_{\phi,k}^{-1}Z_{\psi,k}^{-2}k^{D-4} g_k^2, \quad u(\tilde{\varrho}) = k^{-D}U_k(Z_{\phi,k}^{-1}k^{D-2}\tilde{\varrho}),
\end{align}
where $\tilde{\varrho} = Z_{\phi,k}^{-1}k^{D-2}\varrho$ (and likewise for the VEV $\varrho_0$) and we have suppressed the indices indicating the scale dependence for simplicity. In the following, we shall work solely with dimensionless quantities, and leave the ``tilde'' implicit. Furthermore, we define the bosonic and fermionic anomalous dimensions in usual fashion, $\eta_{\phi,k} = -\partial_t Z_{\phi,k}/Z_{\phi,k}$ and $\eta_{\psi,k} = -\partial_t Z_{\psi,k}/Z_{\psi,k}$. The FRG flow equations can be derived by inserting the ansatz \eqref{eq:LPAprimeansatz} into the Wetterich equation \eqref{eq:Wetterich} and comparing coefficients. In particular, evaluating for constant $\phi_a = (0,0,\sqrt{2\varrho})$ yields the flow equation for the effective potential
\begin{align} \label{eq:FRGflow-potential}
    \partial_t u(\varrho) &= -D u(\varrho) + (D - 2 +  \eta_\phi) \varrho u'(\varrho) \nonumber\\ &\phantom{{}=} {}+ 2 v_D \ell^{\mathrm{(B)},D}_0(u'(\varrho) + 2\varrho u''(\varrho);\eta_\phi) \nonumber\\
    &\phantom{{}=} {}+4 v_D \ell^{\mathrm{(B)},D}_0(u'(\varrho);\eta_\phi) \nonumber\\
    &\phantom{{}=} {} - 4 v_D \left[\tfrac{2N}{3} \ell^{\mathrm{(F)},D}_0(2\varrho g^2; \eta_\psi) + \tfrac{N}{3}\ell^{\mathrm{(F)},D}_0(0; \eta_\psi)\right].
\end{align}
The factor $v_D \coloneq [2^{D+1} \pi^{D/2}\Gamma(D/2)]^{-1}$ arises from integration over the surface of the sphere in $D$-dimensional Fourier space. The threshold functions $\ell_0^{\mathrm{(B)},D}$ and $\ell_0^{\mathrm{(F)},D}$ involve the remaining radial integration and encode the details of the regularization scheme, see Ref.~\cite{bergesreview} for formal definitions.
While the first line of Eq.~\eqref{eq:FRGflow-potential} represents the tree-level flow, the second and third line arise from the fluctuations of the one Higgs mode with mass $2\varrho u''(\varrho)$ and the two Goldstone modes respectively, in full agreement with the Gross-Neveu-SU(2) case~\cite{janssen14}. In the fermion bubble contribution (last line), the first term corresponds to the $2N/3$ gapped modes with mass $2\varrho g^2$, and the second term to the $N/3$ modes that remain gapless in the presence of a constant background $\varrho$.

The definition of the Yukawa coupling is actually ambiguous in the SSB regime, as in general the fermions couple differently to Higgs and Goldstone modes. Assuming the coupling to the Goldstone modes (due to their masslessness) to be the one primarily important for critical behavior \cite{janssen12,janssen14}, we determine the flow of the Yukawa coupling by projecting onto $\phi_1 \bar{\psi} (L_1 \otimes \mathds{1}_2) \psi$ and obtain
\begin{align} \label{eq:FRGflow-Yukawa}
    \partial_t g^2 &= (D - 4 + \eta_\phi + 2 \eta_\psi) g^2 \nonumber\\
    &\phantom{{}=} {}+ 8v_D \ell^{\mathrm{(FB)},D}_{11}(2\varrho_0 g^2, u'_0;\eta_\psi,\eta_\phi) g^4 \nonumber\\
    &\phantom{{}=} {}- 16v_D \varrho_0 u''_0 \ell^{\mathrm{(FBB)},D}_{111}(2\varrho_0 g^2, u'_0, u'_0 + 2\varrho_0 u''_0;\eta_\psi,\eta_\phi) g^4.
\end{align}
Likewise, comparison of coefficients for the kinetic terms gives the anomalous dimensions,
\begin{align}
    \eta_\phi &= \frac{32 N v_D}{3 D} m^{\mathrm{(F)},D}_4(2\varrho_0 g^2;\eta_\psi) g^2 \nonumber\\
    & \phantom{{}={}} {}+ \frac{16v_D}{D} m_{22}^{(\text{B}),D}(u'_0,u'_0+2\varrho_0 u''_0;\eta_\phi)\varrho_0\kern.1em {u''_0}^2\\
    \eta_\psi &= \frac{16 v_D}{3D}\left[m^{\mathrm{(FB)},D}_{12}(0,u'_0;\eta_\psi,\eta_\phi) \right. \nonumber\\
    & \phantom{{}={}} \left.{}+ m^{\mathrm{(FB)},D}_{12}(2\varrho_0 g^2,u'_0;\eta_\psi,\eta_\phi) \right.\nonumber\\
    & \phantom{{}={}} \left.{}+ m^{\mathrm{(FB)},D}_{12}(2\varrho_0 g^2,u'_0 + 2\varrho_0 u''_0;\eta_\psi,\eta_\phi)\right]g^2.
    \label{eq:FRGflow-anomdim}
\end{align}
Here, $\ell_{11}^{\mathrm{(FB)},D}$, $m_{4}^{\mathrm{(F)},D}$, $m_{22}^{(\text{B}),D}$ and $m_{12}^{\mathrm{(FB)},D}$ are further threshold functions defined in Ref.~\cite{bergesreview}.

In this work, we use a linear cutoff, which satisfies an optimization criterion~\cite{litim01}, as well as a sharp cutoff~\cite{janssen12} for comparison. For these regulators, the threshold functions are known analytically, see, e.g., the appendix of Ref.~\cite{janssen12} for an overview.

As a consistency check, we derive, in the limit of small $\epsilon = 4 - D$, the one-loop flow equations of Ref.~\cite{seifert20}. Since the latter employed Wilsonian RG with a sharp cutoff, we need to insert \footnote{The flow equations are non-universal (i.e., dependent on cutoff scheme), so care must be taken when comparing results for fixed-point couplings, and flow equations \emph{a fortiori}, obtained using different methods (this would be true even if we could employ the respective methods exactly). The exponents, on the other hand, are universal and hence scheme-independent. In fact, the loop expansion near the upper critical dimension is universal order by order. We have checked explicitly that the exponents using the linear cutoff agree to $\mathcal{O}(\epsilon)$ with the sharp cutoff result.} the threshold functions corresponding to the sharp cutoff in the flow equations \eqref{eq:FRGflow-potential}--\eqref{eq:FRGflow-anomdim}. We assume the fixed-point effective potential lies in the symmetric regime. Assuming furthermore that fixed-point couplings $g^2_\star = \mathcal{O}(\epsilon)$, $\lambda_{n,\star} = \mathcal{O}(\epsilon^{n-1})$ are parametrically small, we may neglect all higher-order couplings $\lambda_{i > 2}$ in the flow of the effective potential above (i.e., we work in LPA4$'$). Thus,
\begin{align}
    \partial_t \lambda_1 & = (-2+\eta_\phi)\lambda_1 - 10 v_D\frac{\lambda_2}{1+\lambda_1}+\frac{16}{3}v_D N g^2\,,\label{eq:FRGlambda1-4-eps} \\
    \partial_t \lambda_2 & = (-\epsilon+2\eta_\phi)\lambda_2+22 v_D\frac{\lambda_2^2}{(1+\lambda_1)^2}-\frac{32}{3}v_D N g^4\,, \label{eq:FRGlambda2-4-eps}
\end{align}
and
\begin{align}  \label{eq:FRGg2-4-eps}
    \partial_t g^2 = (-\epsilon + \eta_\phi + 2 \eta_\psi) g^2 + 8v_D\frac{g^4}{1+\lambda_1}\,,
\end{align}
with $\eta_\phi = \frac{32}{3} \frac{v_D}{D} N g^2$ and $\eta_\psi = 16\frac{v_D}{D} g^2/(1+\lambda_1)^2$.
We then rescale the couplings $\lambda_2 \to \lambda_2/(4v_D)$ and $g^2 \to g^2/(4v_D)$ and take into account that $v_D=\frac{1}{32\pi^2}+\mathcal{O}(\epsilon)$.
Upon identifying $\lambda_1 \equiv m^2$ and $\lambda_2 \equiv 4!\lambda$, Eqs.~\eqref{eq:FRGlambda1-4-eps}--\eqref{eq:FRGg2-4-eps} coincide precisely with the one-loop flow equations given in Ref.~\cite{seifert20}.

A fixed point of the FRG flow equations is given by $\partial_t g^2 = 0$ and $\partial_t u(\varrho) = 0$ for all $\varrho > 0$. Employing the polynomial expansion of the average potential yields $(n+2)/2$ coupled nonlinear equations for the $(n+2)/2$ couplings $(g^2,\lambda_1,\dots,\lambda_{n/2})$ or $(g^2,\varrho_0,\hat\lambda_2,\ldots,\hat{\lambda}_{n/2})$ depending on regime.
In arbitrary fixed space-time dimension $2<D<4$, these equations can be solved iteratively~\cite{janssen14}.
In $D=3$, we always find a unique fixed point that is characterized by a single relevant direction in the renormalization group sense. Upon increasing the dimension towards $D \nearrow 4$, this fixed point is adiabatically connected to the infrared stable fixed point of the one-loop flow in the $4-\epsilon$ expansion.
We discuss the corresponding critical exponents in the following section, together with the results of the other two approaches.

\section{Discussion}\label{sec:discussion}

\begin{figure*}[tbp]
    \includegraphics[width=0.32\textwidth]{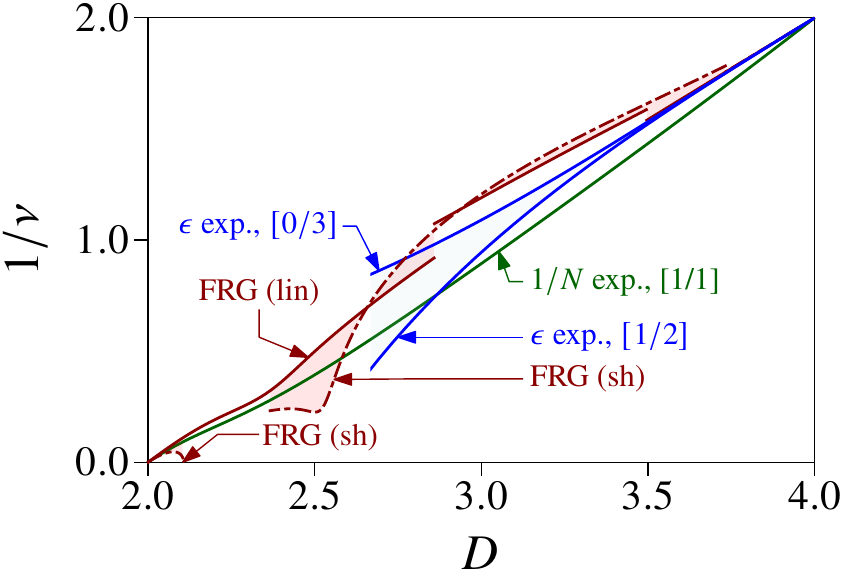}\hfill
    \includegraphics[width=0.32\textwidth]{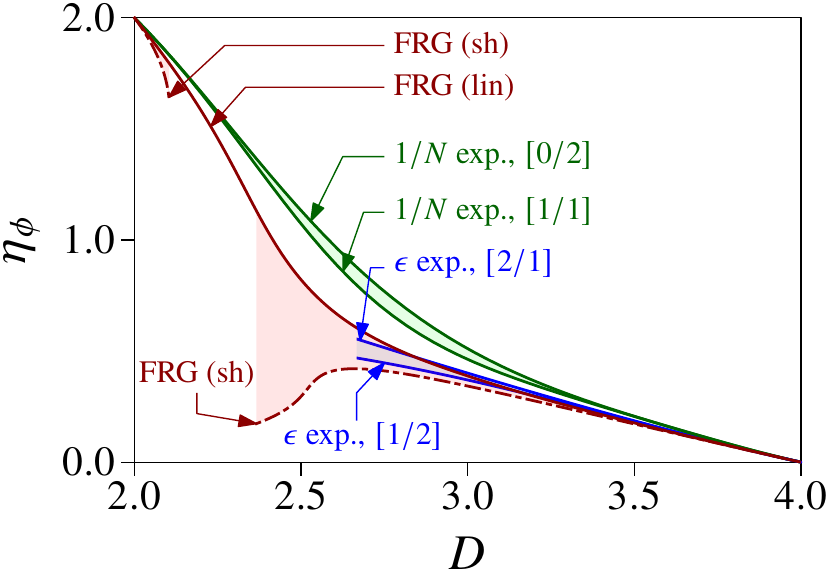}\hfill
    \includegraphics[width=0.32\textwidth]{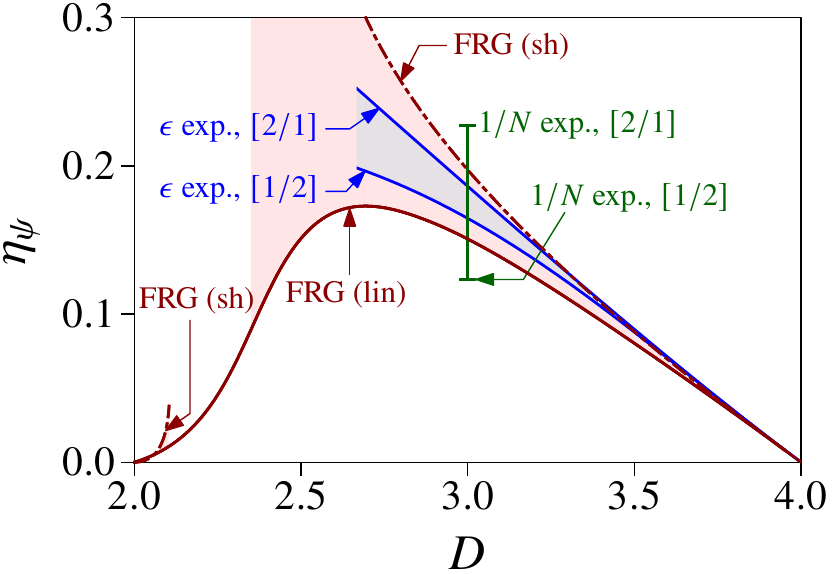}
    \caption{Critical exponents of the Gross-Neveu-SO(3) universality class as a function of space-time dimension $D$ for $N=3$ flavors of two-component Dirac fermions from three-loop $4-\epsilon$ expansion, second-order $1/N$ expansion (third-order for $\eta_\psi$), and FRG in LPA16$'$ using linear (lin) and sharp (sh) regulators.
    $[m/n]$ correspond to different Pad\'e approximants.}
    \label{fig:exponents-vs-D}
\end{figure*}

\begin{figure*}[tbp]
    \includegraphics[width=0.32\textwidth]{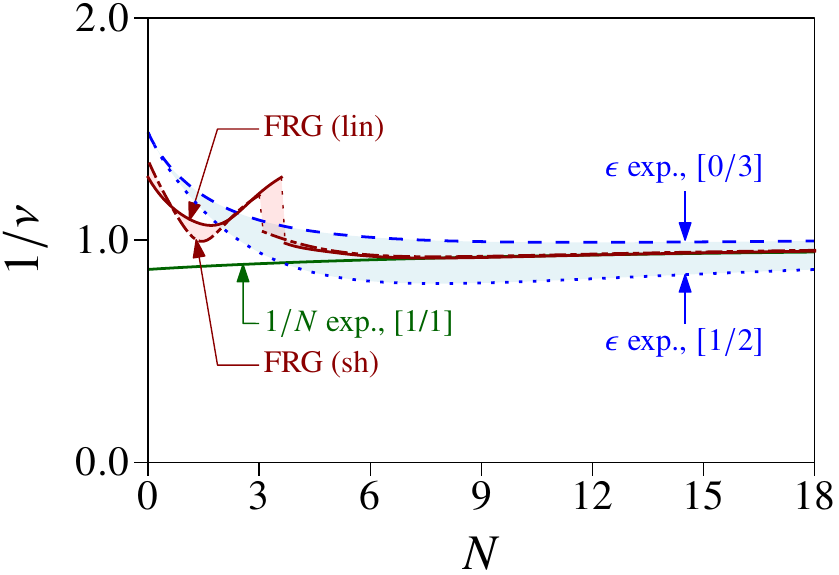}\hfill
    \includegraphics[width=0.32\textwidth]{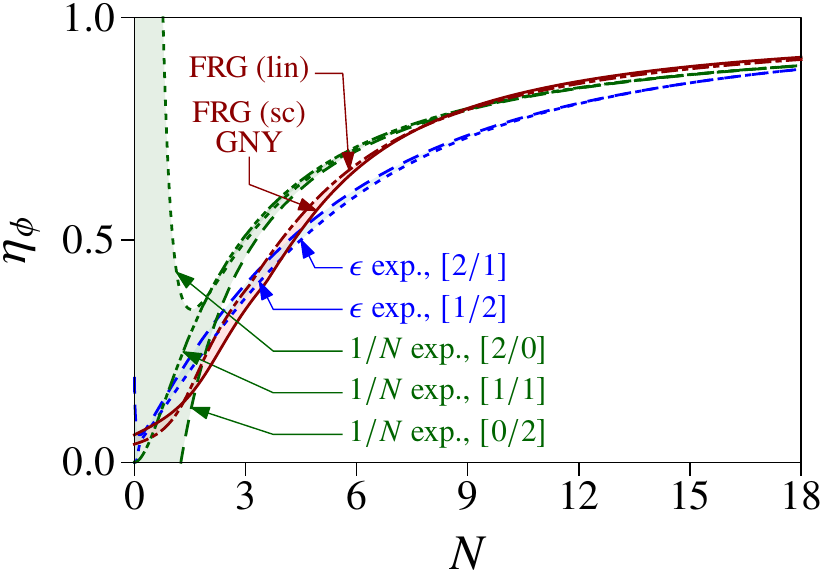}\hfill
    \includegraphics[width=0.32\textwidth]{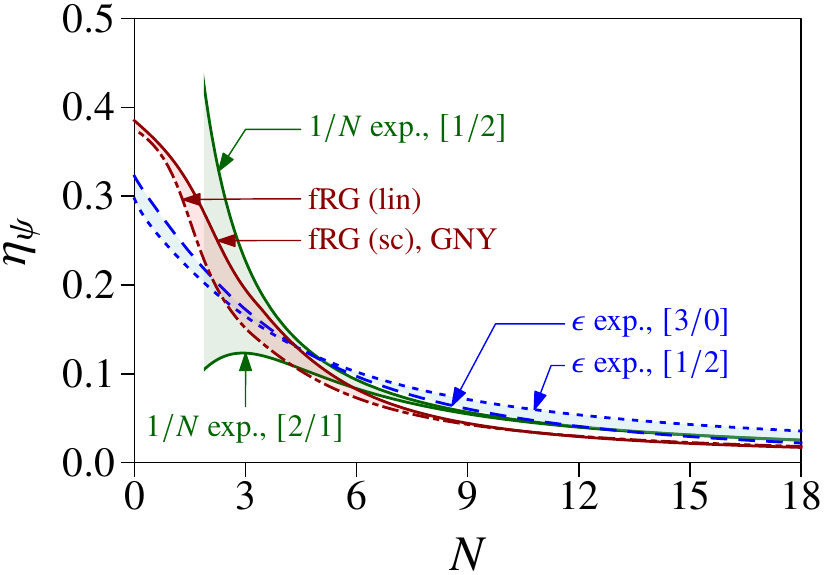}
    \caption{Same as Fig.~\ref{fig:exponents-vs-D}, but as a function of two-component Dirac fermion flavors $N$ in fixed space-time dimension $D = 3$.}
    \label{fig:exponents-vs-N}
\end{figure*}

The quantum critical point is characterized by a set of universal exponents. In this work, we focus on the leading exponents $\nu$ and $\eta_\phi$, as well as the fermion anomalous dimension $\eta_\psi$.
Here, the exponent $\nu$ determines the divergence of the correlation length $\xi$ upon approaching the quantum critical point, while the boson and fermion anomalous dimensions $\eta_\phi$ and $\eta_\psi$ govern the scaling forms of the respective correlators.
We emphasize that the fermionic correlator is not gauge invariant in the spin-orbital models and therefore $\eta_\psi$ does not correspond to an observable quantity in this setting.
However, as the Gross-Neveu-SO(3) universality may in principle also be realized in a model of interacting fermions, in which case $\eta_\psi$ \emph{is} measurable, we also discuss this quantity here.
Subleading quantities that control the corrections to scaling upon approaching the quantum critical point, such as $\omega$, can in principle also be computed within our approaches, but are left for future work.

\begin{table}[tbp]
    \renewcommand{\arraystretch}{1.25}
    \caption{Critical exponents for the Gross-Neveu-SO(3) universality class for $N = 3$ flavors of two-component fermions in $D = 2+1$ space-time dimensions as relevant for the spin-orbital model on the honeycomb lattice~\cite{seifert20} from three-loop $4 - \epsilon$ expansion, second-order $1/N$ expansion (third-order for $\eta_\psi$), and functional renormalization group.
    $[m/n]$ correspond to different Pad\'e approximants.
    For the $(4-\epsilon)$-expansion results ($1/N$-expansion results), we have refrained from showing approximants that exhibit a singularity in $D \in (2,4)$ [in $N \in (0,\infty)$], marked with ``sing.''; those that do not exist are marked ``n.-e.''. A dash (---) signifies that the approximant either entails the computation of terms which go beyond the scope of this work, or conversely does not exhaust all the terms computed in the preceding sections. To obtain the FRG results in LPA$'$, we have treated the bosonic effective potential using a Taylor expansion [i.e., LPA$n'$, with $n \leq 16$ (28) for the linear (sharp) regulator; the error bars correspond to the uncertainty in extrapolating to $n\to\infty$] as well as a pseudospectral decomposition in terms of Chebyshev polynomials.}
    \label{tab:exponents-N-3}
    \begin{tabular*}{\linewidth}{@{\extracolsep{\fill} } l l l c c c}
    \hline\hline
        \multicolumn{2}{l}{$N=3$} & & $1/\nu$ & $\eta_\phi$ & $\eta_\psi$\\
    \hline
        \multicolumn{2}{l}{$4-\epsilon$ expansion}
        & na\"ive & $0.97516$ & $0.39181$ & $0.17234$ \\
        && $[1/2]$ & $0.94472$ & $0.40086$ & $0.16458$ \\
        && $[2/1]$ & sing. & $0.36989$ & $0.18622$ \\
        && $[0/3]$ & $1.09000$ & n.-e. & n.-e. \\
        \multicolumn{2}{l}{$1/N$ expansion}
        & na\"ive & $2.67318$ & $0.49833$ & --- \\
        && $[1/1]$ & $0.89397$ & $0.46276$ & --- \\
        && $[0/2]$ & sing. & $0.51074$ & n.-e. \\
        && na\"ive & --- & --- & $0.22116$ \\
        && $[1/2]$ & --- & --- & $0.12337$ \\
        && $[2/1]$ & --- & --- & $0.22716$ \\
        && $[0/3]$ & --- & --- & n.-e. \\
        FRG & Taylor
        & linear & $1.1901(10)$ & $0.38781(6)$ & $0.15068(8)$ \\
        && sharp & $1.209(4)$ & $0.3434(5)$ & $0.1966(6)$ \\
        & pseudospectral
        & linear & 1.18974 & $0.38781$ & $0.15072$ \\
        && sharp & 1.20465 & $0.34340$ & $0.19649$ \\
    \hline\hline
    \end{tabular*}
\end{table}

\begin{table}[tbp]
    \renewcommand{\arraystretch}{1.25}
    \caption{Same as Table~\ref{tab:exponents-N-3}, but for $N=6$.}
    \label{tab:exponents-N-6}
    \begin{tabular*}{\linewidth}{@{\extracolsep{\fill} } l l l c c c}
    \hline\hline
        \multicolumn{2}{l}{$N=6$} & & $1/\nu$ & $\eta_\phi$ & $\eta_\psi$\\
    \hline
        \multicolumn{2}{l}{$4-\epsilon$ expansion}
        & na\"ive & $0.86069$ & $0.61414$ & $0.09720$\\
        && $[1/2]$ & $0.81514$ & $0.60023$ & $0.10216$ \\
        && $[2/1]$ & $0.96700$ & $0.61484$ & $0.12551$ \\
        && $[0/3]$ & $1.01291$ & n.-e. & n.-e. \\
        \multicolumn{2}{l}{$1/N$ expansion}
        & na\"ive & $1.28320$ & $0.70572$ & --- \\
        && $[1/1]$ & $0.91136$ & $0.70076$ & --- \\
        && $[0/2]$ & $1.26614$ & $0.71005$ & n.-e. \\
        && na\"ive & --- & --- & $0.09275$\\
        && $[1/2]$ & --- & --- & $0.08341$ \\
        && $[2/1]$ & --- & --- & $0.09317$ \\
        && $[0/3]$ & --- & --- & n.-e. \\
        FRG & Taylor 
        & linear & 0.9294(6) & 0.66947(6) & 0.073170(17)\\
        && sharp & 0.926(3) & 0.6598(4) & 0.08257(16)\\
        & pseudospectral 
        & linear & 0.92961 & 0.66948 & 0.073165\\
        && sharp & 0.93245 & 0.65980 & 0.082570\\
    \hline\hline
    \end{tabular*}
\end{table}

\begin{table}[tbp]
    \renewcommand{\arraystretch}{1.25}
    \caption{Same as Table~\ref{tab:exponents-N-3}, but for $N=12$. For the FRG results, we have omitted the error bars corresponding to the uncertainty in the extrapolation of the Taylor expansion of the effective potential, as they are smaller than $2\times10^{-5}$.}
    \label{tab:exponents-N-12}
    \begin{tabular*}{\linewidth}{@{\extracolsep{\fill} } l l l c c c}
    \hline\hline
        \multicolumn{2}{l}{$N=12$} & & $1/\nu$ & $\eta_\phi$ & $\eta_\psi$\\
    \hline
        \multicolumn{2}{l}{$4-\epsilon$ expansion}
        & na\"ive & $0.84820$ & $0.80614$ & $0.04095$ \\
        && $[1/2]$ & $0.82616$ & $0.80659$ & $0.05391$ \\
        && $[2/1]$  & $0.91427$ & $0.80775$ & sing. \\
        && $[0/3]$ & $0.99001$ & n.-e. & n.-e. \\
        \multicolumn{2}{l}{$1/N$ expansion}
        & na\"ive & $1.00325$ & $0.84199$ & --- \\
        && $[1/1]$  & $0.93326$ & $0.84134$ & --- \\
        && $[0/2]$ & $0.98522$ & $0.84280$ & n.-e. \\
        && na\"ive & --- & --- & $0.04054$\\
        && $[1/2]$ & --- & --- & $0.03995$\\
        && $[2/1]$ & --- & --- & $0.04056$ \\
        && $[0/3]$ & --- & --- & n.-e. \\
        FRG & Taylor 
        & linear & 
        0.93660 & 0.85180 & 0.02992 \\
        && sharp & 0.93282 & 0.85700 & 0.02941 \\
        & pseudospectral 
        & linear & 0.93660  & 0.85180  & 0.02992\\
        && sharp & 0.93282  & 0.85700  & 0.02941\\
    \hline\hline
    \end{tabular*}
\end{table}

Figure~\ref{fig:exponents-vs-D} shows our results for $1/\nu$, $\eta_\phi$, and $\eta_\psi$ as a function of space-time dimension $2<D<4$ for $N=3$ flavors of two-component Dirac fermions, which is the case relevant for the spin-orbital models.
Since the $4-\epsilon$ and large-$N$ expansions are \emph{per se} only valid asymptotically for vanishing expansion parameter, we have employed different Pad\'e approximants, marked as ``$[m/n]$'' with integer $m$ and $n$ in the plots (note that $n = 0$ simply corresponds to the na\"ive extrapolation of the series expansion to finite $\epsilon$ or $1/N$). The difference between the different Pad\'e approximants provides a simple estimate for the systematic error of the extrapolation to finite $\epsilon$ and $1/N$, respectively.
For the same purpose, in the FRG calculation, we have applied two different regularization schemes, marked as ``lin'' for the linear cutoff and ``sh'' for the sharp cutoff.
We note that in the sharp-cutoff scheme, there is no stable fixed point for $2.104 < D < 2.366$ as a consequence of fixed-point collisions at the lower and upper bound of this interval. In this cutoff scheme, the fixed point in $D = 2 + \epsilon$ dimensions for small $\epsilon$ is therefore \emph{not} adiabatically connected to the fixed point in $D = 4- \epsilon$ dimensions.
We also note that in both cutoff schemes, the FRG fixed point for $N=3$ is located in the symmetric regime for $D=2 + \epsilon$ and $D=4 - \epsilon$ for small $\epsilon$, but in the symmetry-broken regime for $D=3$. This leads to discontinuities in $1/\nu$ at those values of $D$, at which the minimum of the fixed-point potential becomes finite, see left panel of Fig.~\ref{fig:exponents-vs-D}.
Reassuringly, we observe that all curves approach each other near the upper critical space-time dimension $D_\mathrm{up}=4$, as it should be~\cite{janssen14}.

Figure~\ref{fig:exponents-vs-N} shows the critical exponents for the physical dimension $D=2+1$ as a function of the flavor number $N$.
For $N$ sufficiently large and increasing, the deviations between the different approaches decrease for increasing $N$ and vanish in the limit $N\to\infty$ as expected.
Note that for large $N$, the fixed point in the FRG calculation is again located in the symmetric regime, in analogy to the behavior of the Gross-Neveu-$\mathbbm{Z}_2$ model~\cite{hoefling02,braun11}. The transition from symmetry-broken to symmetric regime upon increasing $N$ is accompanied by a jump in $1/\nu$, similar to the transition as a function of $D$ discussed above.

The numerical estimates for the physical dimension $D=2+1$ are given in Table~\ref{tab:exponents-N-3} for $N=3$ and in Tables~\ref{tab:exponents-N-6}--\ref{tab:exponents-N-12} for larger values of $N$.
{Note that some Pad\'e approximants develop unphysical poles as a function of the expansion parameter, which render them unreliable as extrapolations of the asymptotic series expansion. The corresponding entries are hence labelled ``sing.'' in the tables. Note also that the maximally asymmetric Pad\'e approximants $[0/n]$ cannot fulfil the boundary conditions needed to extrapolate $\eta_\psi$ in both expansions, as well as $\eta_\phi$ in the $(4-\epsilon)$-expansion. Such non-existent approximants are marked as ``n.-e.'' in the tables.}
Overall, we observe a fair agreement of the estimates from the three different approaches. 
In order to obtain final estimates for the three exponents from the combination of the three different approaches we first average over the values of the different approximants and regularization schemes, respectively, within a given approach. {Thus, for both the $4-\epsilon$ and large-$N$ expansions, we average over all well-behaved Pad\'e approximants. The na\"ive extrapolations, which formally constitute $[m/0]$-type Pad\'e approximants, are included in the respective average if and only if they are sandwiched by two well-behaved ``proper'' Pad\'e approximants $[m_1/n_1], [m_2/n_2]$ with $n_1,n_2 \neq 0$. For the FRG calculation, we average first between the Taylor expansion and pseudospectral decomposition results for a given regulator, and then average over the two regulators. The last step is to average over the three methods, which then yields our final best-guess estimates.}
The spread of the three mean values provides a rough estimate for the accuracy of our final result.
{We emphasize that this procedure may potentially underestimate the systematic error involved in the different calculations and should therefore only be understood simply as a measure of consistency of the three approaches.}
For the physically relevant case of $N = 3$ flavors of two-component Dirac fermions in $D=2+1$ space-time dimensions~\cite{seifert20}, we obtain the critical exponents as
\begin{align} \label{eq:exponents-N=3}
    &N=3:& 
    1/\nu & = 1.03(15), &
    \eta_\phi & = 0.42(7), &
    \eta_\psi & = 0.180(10).
\end{align}
Equation~\eqref{eq:exponents-N=3} represents the main result of this work.
As there appears to be no dangerously irrelevant coupling in the theory, we expect hyperscaling to be satisfied. The critical exponents $\alpha$, $\beta$, $\gamma$, and $\delta$ can then be obtained from $\nu$ and $\eta_\phi$ with the help of the usual hyperscaling relations~\cite{herbutbook}.
For completeness, we also quote the estimates obtained for larger values of $N$, which may be relevant for models with microscopic fermionic degrees of freedom,
\begin{align} \label{eq:exponents-N=6}
    &N=6:&
    1/\nu & = 0.98(10), &
    \eta_\phi & = 0.66(5), \nonumber\\ &&
    \eta_\psi & = 0.094(18),
\end{align}
and
\begin{align}
    &N=12:& 
    1/\nu & = 0.93(4), &
    \eta_\phi & = 0.83(4), \nonumber\\ &&
    \eta_\psi & = 0.041(12).
\end{align}
%

\section{Summary and outlook}\label{sec:conclusions}

In this work, we have investigated the critical behavior of the $(2+1)$-dimensional Gross-Neveu-SO(3) universality class in terms of the universal critical exponents $\nu$, $\eta_\phi$, and $\eta_\psi$ by means of different sophisticated field-theoretical techniques.
The fractionalized counterpart of the Gross-Neveu-SO(3) universality class, dubbed Gross-Neveu-SO(3)*, may be realized in spin-orbital magnets with strong exchange frustration~\cite{seifert20}. In contrast to the fractionalized bosonic universality classes~\cite{isakov12}, in the fractionalized fermionic universality classes, not only the correlation-length exponent $\nu$, but also the order-parameter anomalous dimension $\eta_\phi$ agrees with the value of the corresponding conventional fermionic universality class.
This allows us to obtain estimates for both Gross-Neveu-SO(3) and Gross-Neveu-SO(3)* from the same calculation.
We emphasize, however, that the fermionic correlator is not gauge invariant in the spin-orbital model. Our estimate for the fermion anomalous dimension $\eta_\psi$ therefore applies only to the conventional Gross-Neveu-SO(3) universality class.

The Gross-Neveu-SO(3) theory is different from the previously studied Gross-Neveu-type models, as it features a symmetry-breaking transition between two semimetallic phases, with only a partial gap opening in the ordered phase.
This leads to values for the critical exponents that strongly differ from those of the semimetal-to-insulator Gross-Neveu transitions~\cite{zerf17}.
In particular, the order-parameter anomalous dimension $\eta_\phi$ in the Gross-Neveu-SO(3) model is significantly smaller than $\eta_\phi$ in any of the other Gross-Neveu-type models for the same number of fermion flavors.
These difference may be readily observable in numerical simulations of suitable lattice models.

For the future, it would be interesting to study further properties of the Gross-Neveu-SO(3) universality class. In particular, it might be worthwhile to examine the finite-size spectrum on the torus, which was recently investigated in the conventional Gross-Neveu-$\mathbbm{Z}_2$ universality class~\cite{schuler21}, both in the conventional Gross-Neveu-SO(3) and the fractionalized Gross-Neveu-SO(3)* cases.

\begin{acknowledgments}

We are grateful to Sreejith Chulliparambil, Xiao-Yu Dong, Urban Seifert, Hong-Hao Tu, and Matthias Vojta for illuminating discussions and collaborations on related topics.
We thank Matthias Steinhauser for correspondence and for providing us with the programs \texttt{q2e} and
\texttt{exp} to carry out the three-loop calculations.
Figures \ref{sde2pt}, \ref{sde3pt}, and \ref{3ptcf} were drawn with the \texttt{axodraw} package \cite{axodraw}.
S.R.\ and L.J.\ acknowledge support by the Deutsche Forschungsgemeinschaft (DFG) through SFB 1143 (project A07, project id 247310070), the W\"{u}rzburg-Dresden Cluster of Excellence {\it ct.qmat} (EXC 2147, project id 390858490), and the Emmy Noether program (JA2306/4-1, project id 411750675).
J.A.G.\ was supported by the DFG through a Mercator Fellowship.
M.M.S. acknowledges support by the DFG through SFB 1238 (projects C02 and C03, project id 277146847). 

\end{acknowledgments}

\appendix

\section{\texorpdfstring{Critical exponents for $\boldsymbol{N\geq3}$ in $\boldsymbol{4-\epsilon}$ expansion}{Critical exponents for N>=3 from 4-epsilon expansion}} \label{app:epsilon}

In this appendix, we give the full expressions for the critical exponents from the $4-\epsilon$ expansion at three-loop order for arbitrary $N\geq 3$. These are also provided electronically in the ancillary file \texttt{GNSO3-exponents.m} of the Supplemental Material~\cite{suppl}.
The file contains the series in $\epsilon \equiv \texttt{eps}$ for the inverse correlation length exponent $\nu^{-1}\equiv \texttt{nuinveps}$ as well as the boson and fermion anomalous dimensions $\eta_\phi\equiv\texttt{etaphieps}$ and $\eta_\psi\equiv\texttt{etapsieps}$, respectively.
Further, we use $N \equiv \texttt{n}$ in the file. 
The full expressions are
\begin{widetext}
\begin{align}
    \frac{1}{\nu}\,=&\, 2-\frac{17N+5 (s+6)}{22(N+6)}\epsilon+\frac{820 N^4-N^3 (820 s-172050)-N^2 (19032 s-65745)-18 N (7417 s-111450)-179280 (s+6)}{10648 (N+6)^3 s}\epsilon^2\nonumber\\
    &+\frac{3}{10307264 (N+6)^5 s^3}\bigg\{130160 N^8
    -40 N^7 (3254 s-792723)-42 N^6 (615046 s-62828375)\nonumber\\
    &+N^5 (38744100900-1283135016 s)-90 N^4 (76274956 s+4047257499)-27 N^3 (650749372 s+61741071045)\nonumber\\
    &+324 N^2 (690975808 s-10686990915)-972 N (39813404 s+2231925285)-32313945600 (s+6)\nonumber\\
    &+88\left(N^3+126 N^2+756 N+216\right) \Big[2960 N^5+N^4 (417300-2960 s)-150 N^3 (1598 s-31167)\nonumber\\
    &-9 N^2 (181016 s-2657505)-162 N (15058 s-43965)+3836160 (s+6)\Big]\zeta_3
    \bigg\}\epsilon^3 + \mathcal{O}(\epsilon^4)\,,\displaybreak[0]\\
    \eta_\phi\, =&\, \frac{N}{N+6}\epsilon + \frac{3 }{968 (N+6)^3}\left[1010 N^2+N (200 s-867)+120 (s+6)\right]\epsilon^2\nonumber\\
    &+\frac{3}{468512 (N+6)^5 s}
    \Big[-44590 N^5+6 N^4 (41594 s-1353735)+15 N^3 (494903 s+858426)+72 N^2 (342319 s-6404430)\nonumber\\
    &+27 N (829517 s+593640)+11171520 (s+6) -3162456 \left(N^2+9 N+18\right) N s \zeta_3
    \Big]\epsilon^3+ \mathcal{O}(\epsilon^4)\,,\displaybreak[0]\\
    \eta_\psi\, =&\,\frac{3}{2 (N+6)}\epsilon+\frac{3}{1936 (N+6)^3}\left[-736 N^2+5 N (2 s+123)+600 s+9045\right]\epsilon^2\nonumber\\
    &-\frac{9}{937024 (N+6)^5 s}\Big[
   1570 N^5+32 N^4 (866 s+13605)+3 N^3 (457279 s+9071670)+N^2 (65284110-55782 s)\nonumber\\
   &+N (330505920-39115413 s)+81 (339000-719473 s)
   +3162456 \left(N^2+9 N+18\right) s \zeta_3 \Big]\epsilon^3+ \mathcal{O}(\epsilon^4)\,,
\end{align}
\end{widetext}
where we have abbreviated $s\coloneqq \sqrt{N^2+120N+36}$.

\section{\texorpdfstring{Critical exponents for $\boldsymbol{2<D<4}$ in $\boldsymbol{1/N}$ expansion}{Critical exponents for 2<D<4 from 1/N expansion}}
\label{app:large-N}

In this appendix, we record the full $D$-dimensional expressions for the various
critical exponents that have been computed in the large-$N$ expansion. These are also provided electronically in the ancillary file \texttt{GNSO3-exponents.m} as Supplemental Material~\cite{suppl}. 
The file contains the series in $1/N \equiv \texttt{1/n}$ for the inverse correlation length exponent $\nu^{-1}\equiv \texttt{nuinvn}$ as well as the boson and fermion anomalous dimensions $\eta_\phi\equiv\texttt{etaphin}$ and $\eta_\psi\equiv\texttt{etapsin}$, respectively.
We denote the numerical coefficients of the $1/N$ series as
\begin{align}
    \chi & = \sum_{n=0}^\infty \chi_n \left(\frac{1}{N}\right)^n, &
    \lambda & =  \sum_{n=0}^\infty \lambda_n \left(\frac{1}{N}\right)^n, &
    \eta_\psi & =  \sum_{n=0}^\infty \eta_n \left(\frac{1}{N}\right)^n, &
\end{align}
The leading-order terms are identical in all Gross-Neveu-like universality classes,
\begin{align}
    \chi_0 & = 0\,, &
    \lambda_0 & = \mu - 1\,, &
    \eta_0 & = 0\,,
\end{align}
where we have abbreviated $\mu \equiv D/2$.
To order $\mathcal O(1/N)$, we recover the expressions that were originally determined in Ref.~\cite{seifert20},
\begin{align}
\chi_1 &= \frac{\mu}{2 (\mu-1)} \eta_1\,, &
\lambda_1 &= - (2 \mu-1) \eta_1\,,
\end{align}
where
\begin{equation}
\eta_1  =  -~
\frac{3 \Gamma(2\mu-1)}{\mu\Gamma(1-\mu)\Gamma(\mu-1)\Gamma^2(\mu)} \,.
\end{equation}
At next order, we have
\begin{align}
\eta_2 &= \left[ \frac{(3 \mu-2)}{2 (\mu-1)} \Psi(\mu)
+ \frac{13 \mu^2-12 \mu+2}{4 \mu (\mu-1)^2} \right] \eta_1^2\,, \\
\chi_2 &= \left[ \frac{\mu (3 \mu-2)}{4 (\mu-1)^2} \Psi(\mu)
- \frac{\mu (4 \mu^2-3 \mu-2)}{4 (\mu-1)^2} 
+ \frac{9 \mu^2 \Theta(\mu)}{8 (\mu-1)} \right] \eta_1^2\,,
\end{align}
for the field anomalous dimensions, while the correction to the exponent
relating to $\nu$ is
\begin{widetext}
\begin{eqnarray}
\lambda_2 &=&
\left\{ \frac{3\mu (\mu^2-2 \mu+4)}{4 (\mu-1) (\mu-2)^2 \eta_1}
- \frac{32 \mu^6-178 \mu^5+349 \mu^4-265 \mu^3-14 \mu^2+128 \mu-32}
{8 (\mu-1)^2 (\mu-2)^2} \Psi(\mu)
\right. \nonumber \\
&& \left. 
{}- \frac{7 \mu^2 (2 \mu-3)}{8 (\mu-1) (\mu-2)}
\left[\Psi^2(\mu) + \Phi(\mu) \right]
- \frac{3\mu^2 (4 \mu^2-27 \mu+28)}{8 (\mu-1) (\mu-2)} \Theta(\mu)
\right. \nonumber \\
&& \left. 
{}+ \frac{64 \mu^8-528 \mu^7+1650 \mu^6-2375 \mu^5+1367 \mu^4+218 \mu^3
- 632 \mu^2+256 \mu-32}{16 \mu (\mu-1)^3 (\mu-2)^2}
\right\} \eta_1^2 \,.
\end{eqnarray}
At this order, derivatives of the Euler $\Gamma$ function arise, which is
apparent in the functions
\begin{align}
\Psi(\mu) &= \psi(2\mu-1)  -  \psi(1)  +  \psi(2-\mu)  -  \psi(\mu)\,,
&
\Theta(\mu) &= \psi^\prime(\mu)  -  \psi^\prime(1)\,,
\end{align}
where $\psi(z) = \mathrm d \ln \Gamma(z)/(\mathrm d z)$ is the Euler $\psi$ function. Finally, the large-$N$ conformal
bootstrap formalism produced
\begin{eqnarray}
\eta_3 &=&
\left\{
\frac{3 (3 \mu-2)^2}{8 (\mu-1)^2} \Psi^2(\mu)
+ \frac{(3 \mu-2)^2}{8 (\mu-1)^2} \Phi(\mu)
- \frac{8 \mu^7-5 \mu^6-8 \mu^5-182 \mu^4+414 \mu^3-288 \mu^2+80 \mu-8}
{16 \mu^2 (\mu-1)^4}
\right. \nonumber \\
&& \left.  
{}- \frac{4 \mu^5-7 \mu^4-101 \mu^3+178 \mu^2-88 \mu+12}
{8 \mu (\mu-1)^3} \Psi(\mu)
+ \frac{3 \mu^3+24 \mu^2+12 \mu-8}{32 (\mu-1)^2}
\left[ \Theta(\mu) + \frac{1}{(\mu-1)^2} \right]
\right. \nonumber \\
&& \left.  
{}+ \frac{9 \mu^2}{8 (\mu-1)} \left[ \Theta(\mu) + \frac{1}{(\mu-1)^2} \right]
\Psi(\mu)
+ \frac{9 \mu^2}{16 (\mu-1)} \Xi(\mu)
\left[ \Theta(\mu) + \frac{1}{(\mu-1)^2} \right] \right\} \eta_1^3\,,
\end{eqnarray}
\end{widetext}
where an additional function $\Xi(\mu)$ appears. It is related to a particular
two-loop self-energy diagram that was defined as $I(\mu)$ in Eq.~(16) of 
Ref.~\cite{vasilev82} and is connected to $\Xi(\mu)$ by
\begin{equation}
I(\mu)  =  - \frac{2}{3(\mu-1)}  +  \Xi(\mu) \,.
\end{equation}
In Ref.~\cite{bgk97} it was shown to be related to derivatives with respect to the
parameter dependence of an ${}_4 F_3$ hypergeometric function and its
$\epsilon$ expansion was given to very high orders near two and four
space-time dimensions. The three-dimensional value was given in Ref.~\cite{vasilev82} as
\begin{equation}
I(\tfrac32)  =  2 \ln 2  -  \frac{21}{\pi^2} \zeta_3 \,.
\end{equation}
%

\section{Convergence of pseudospectral and Taylor expansions}
\label{app:FRGanom}

\begin{figure}[tbp]
    \includegraphics[width=\columnwidth]{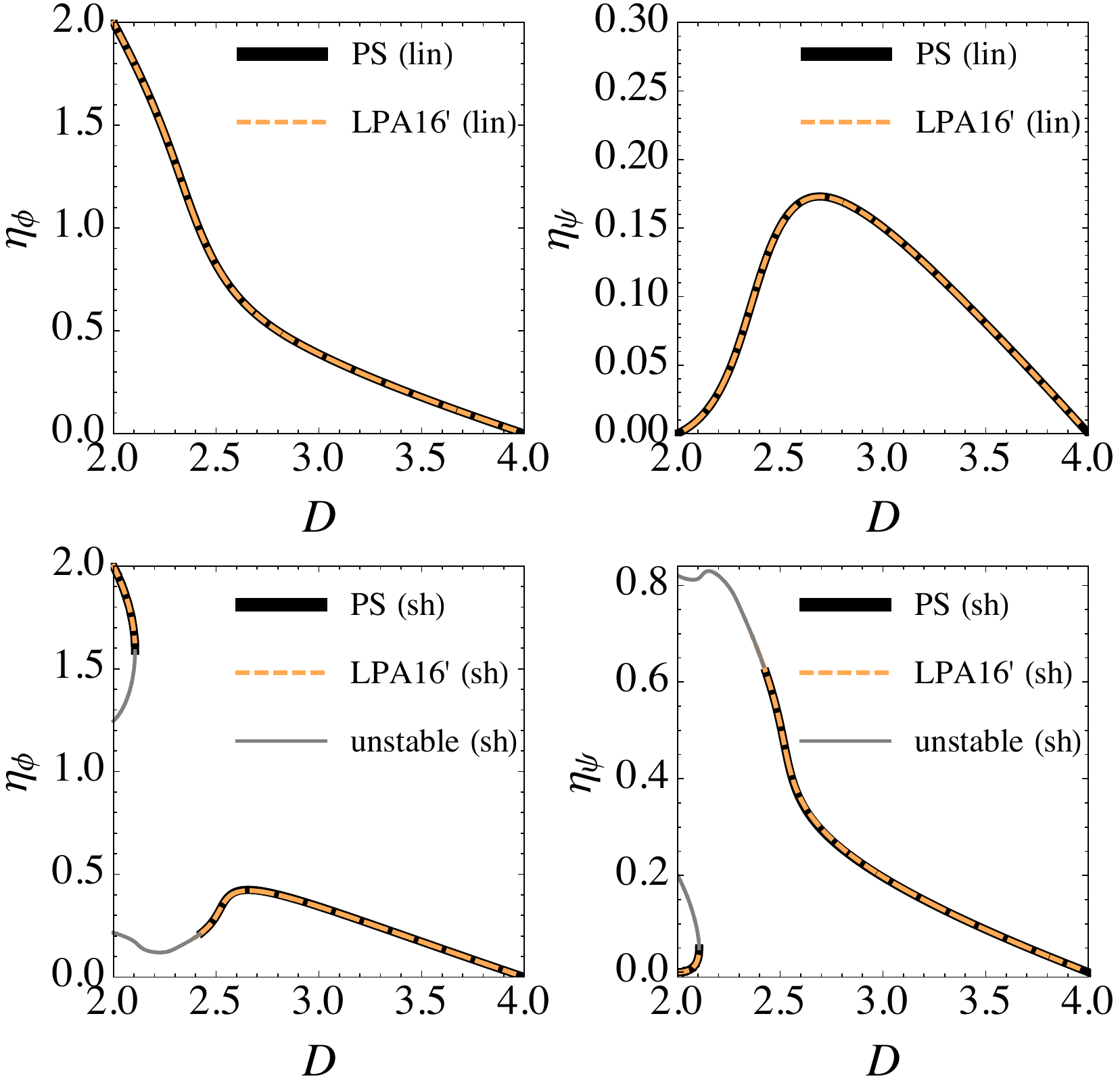}
    \caption{Comparison of the results for the boson and fermion anomalous dimensions obtained from the FRG using both the Taylor expansion in LPA16$'$, and the pseudospectral decomposition with $n_T=n_R=9$ collocation points. The curves match almost perfectly in the whole range $2 <D< 4$.
    Top panels: Linear (lin) cutoff. 
    Bottom panels: Sharp (sh) cutoff. 
    In the sharp cutoff scheme, there is no stable fixed point between $2.104 < D < 2.366$.}
    \label{fig:FRGanom}
\end{figure}

In Sec.~\ref{sec:FRGmethod}, we have introduced two different expansion schemes to find approximate solutions for the FRG flow and the fixed points of the effective potential, i.e., a finite-order Taylor expansion and an expansion based on a pseudospectral decomposition using Chebyshev polynomials. While the advantage of the Taylor expansion is its simple implementation which has proven to work well for many purposes, the Chebyshev expansion may provide superior convergence properties in some cases, e.g., going towards two dimensions (see the discussion in the main text).
To check the reliability of our FRG calculations, we directly compare the results from the Taylor expansion and the pseudospectral expansion for the anomalous dimensions for $2<D<4$, see Fig.~\ref{fig:FRGanom}.
We find excellent agreement between the results from the Taylor expansion and the pseudospectral methods in the whole range of dimensions between two and four.

%

\end{document}